# Charge-Unified Semiconductor Switching Theory

Wucheng Ying, Jinwei Qi, Hui Zhao, Feng An, Ziheng Yu, Yanfeng Shen, C.Q. Jiang, Yanghong Xia, Ameer Janabi, Biao Zhao, Stephan Goetz, Frede Blaabjerg and Teng Long

**Abstract**

Semiconductors and downstream applications sustain the electronic, information, energy and industrial systems underpinning society. Their sustainability is an urgent global priority, particularly as global electricity generation is projected to rise more than 2.5-fold by 2050[1]. However, since 1947[2], a unified, global view of circuit elements as media for charge redistribution and transfer, which reveals switching inertia and dynamical nature of switching, and connects microscopic and macroscopic domains across semiconductor value chain within a unified theoretical language has yet to be established[3]; switching lacks a unified mechanistic account of its physical origins and spatiotemporal evolution, with fundamental disconnects between charge- and energy-conservation frameworks, across carrier dynamics and across equivalent-circuit formalisms. These limitations fragment the domains and impede sustainability gains, particularly those requiring cross-domain causal information. Here we present Charge-Unified Semiconductor Switching Theory (CUSST), a general theory that unifies circuit elements through a charge-mediated view, reveals inertia and the dynamical nature of switching, achieves aforementioned long-missing unifications, and establishes a unified formulation across conceptual, mechanistic, formal and analytical levels, with unprecedented simplicity through unifications. CUSST establishes a unified micro–macro spatiotemporal view of switching, generalizes circuit theory, extends conservation laws and guides theoretical systems. For example, a CUSST-based switching-energy-loss model (errors: 0.88–11.60%) achieves a 17-fold average error reduction compared to the conventional model (errors: 34.41–80.05%); CUSST enables unprecedented causal–mechanistic interpretability to reveal the underlying switching dynamics across scenarios. More broadly, CUSST may support domains along the semiconductor value chain, such as semiconductor materials[4], chip design[5], packaging[6], reliability[7,8], thermal engineering[9,10] and downstream applications such as power electronics and potentially extend to broader systems such as computing, communications and integrated circuits.[11,12] By offering cross-domain causal information, CUSST could advance these fragmented domains towards a unified discipline including through enabling full-chain causal feedback and cross-domain co-design and research, and unlock sustainability gains beyond individual-domain optimization.

## Nomenclature

Table 1| NOMENCLATURE

| Symbol | Unit | Definition |
|---|---|---|
| $S_{\mathrm{x}}$ | N/A | Switch index, e.g., $S_1$ denotes the upper switch and $S_2$ denotes the lower switch. |
| $R_{\mathrm{Sx}}$ | Ω | Equivalent resistance of $S_{\mathrm{x}}$. |
| $v_{\mathrm{ds,Sx}}$ | V | Drain-source voltage of $S_{\mathrm{x}}$. |
| $v_{\mathrm{GG}}$ | V | Output voltage of the gate driver of $S_1$. |
| $v_{\mathrm{gs,Sx}}$ | V | Gate-source voltage of $S_{\mathrm{x}}$. |
| $V_{\mathrm{th,Sx}}$ | V | Threshold voltage of $S_{\mathrm{x}}$. |
| $V_{\mathrm{gp}}$ | V | Gate-source voltage at the Miller plateau (assumed constant) |
| $C_{\mathrm{oss,Sx}}$ | pF | Output capacitance of $S_{\mathrm{x}}$. |
| $C_{\mathrm{Sx}}$ | pF | Overall equivalent capacitance of $S_{\mathrm{x}}$. |
| $C_{\mathrm{gs,Sx}}$ | pF | Gate-source capacitance of $S_{\mathrm{x}}$. |
| $C_{\mathrm{gd,Sx}}$ | pF | Gate-drain capacitance (also known as Miller capacitance) of $S_{\mathrm{x}}$. |
| $C_{\mathrm{ds,Sx}}$ | pF | Drain-source capacitance of $S_{\mathrm{x}}$. |
| $C_{\mathrm{x}}$ | pF | A generic linear or nonlinear capacitance. |
| $i_{\mathrm{d,Sx}}$ | A | The drain current of $S_{\mathrm{x}}$. |
| $i_{\mathrm{d,Sx}}$ | A | The channel current of $S_{\mathrm{x}}$. |
| $i_{\mathrm{g,Sx}}$ | A | Gate driving current of $S_{\mathrm{x}}$. |
| $i_{\mathrm{L}}$ | A | Load current. |
| $i_{\mathrm{RSx}}$ | A | Instantaneous current through $R_{\mathrm{Sx}}$. |
| $i_{\mathrm{DC}}$ | A | DC-source current. |
| $i_{\mathrm{Cgs,Sx}}$ | A | Displacement current of $C_{\mathrm{gs,Sx}}$. |
| $i_{\mathrm{Cgd,Sx}}$ | A | Displacement current of $C_{\mathrm{gd,Sx}}$. |
| $i_{\mathrm{Cds,Sx}}$ | A | Displacement current of $C_{\mathrm{ds,Sx}}$. |
| $i_{\mathrm{C,Sx}}$ | A | Displacement current of $C_{\mathrm{Sx}}$. |
| $i_{\mathrm{RSx}}$ | A | Conductive current through the equivalent capacitance $R_{\mathrm{Sx}}$. |
| $i_{\mathrm{inj}}$ | A | Injection current. |
| $i_{\mathrm{ext}}$ | A | Extraction current. |
| $i_{\mathrm{rec}}$ | A | Equivalent recombination current. |
| $V_{\mathrm{DC}}$ | V | DC-link voltage. |
| $E_{\mathrm{on}}$ | μJ | The turn-on energy dissipation. |
| $E_{\mathrm{gd,Sx}}(v)$ | μJ | The energy stored in $C_{\mathrm{gd,Sx}}$ at drain-gate voltage of v. |
| $E_{\mathrm{ds,Sx}}(v)$ | μJ | The energy stored in $C_{\mathrm{ds,Sx}}$ at drain-source voltage of $v$. |
| $E_{\mathrm{oss,Sx}}(v)$ | μJ | The energy stored in the output capacitance of $S_x$ at drain-source voltage of $v$. |
| $Q_{\mathrm{oss,Sx}}(v)$ | nC | The charge stored in the output capacitance of $S_x$ at drain-source voltage of $v$. |
| $Q_{\mathrm{x}}(v)$ | nC | The charge stored in a generic capacitance at drain-source voltage of $v$. |
| $\Delta Q_{\mathrm{x}}$ | nC | The change in stored charge. |
| $e_{\mathrm{CUSST\text{-}based}}$ | N/A | Error of the CUSST-based model's prediction results w.r.t. the measured results. |
| $e_{\mathrm{con}}$ | N/A | Error of the conventional model's prediction results w.r.t. the measured results. |
| HS | N/A | Hard switching; turn-on without prior $v_{\mathrm{ds}}$ discharge. |
| iZVS | N/A | Incomplete zero-voltage switching; turn-on with partial $v_{ds}$ discharge and residual $v_{ds}$. |
| ZVS | N/A | Zero-voltage switching; turn-on after $v_{ds}$ is discharged to approximately zero. |

## Introduction

The International Energy Agency and World Bank project that AI and robotics-driven intelligentization[13,14], transport electrification[15], heat pumps[16], hydrogen production[17], digitalization[14] and advanced manufacturing[13] are driving unprecedented electricity demands, with global generation exceeding 2.5 times today's level by 2050[1]. Greener semiconductors downstream applications are therefore an increasingly urgent priority for global sustainability. By 2030, semiconductor-enabled power electronics could process over 80% of U.S. electricity,[18] while global data centres could emit 399 million tonnes of $CO_2$ and consume 9.3 trillion litres of water annually—enough to meet the domestic water needs of 1.3 billion people in sub-Saharan Africa[19]. Yet semiconductor losses remain a major source of energy inefficiency and water-cooling demand. For example, in AI data centres, power-electronics losses can reach about 12% of total facility power[20], with semiconductor loss forming a major contribution, while graphics processing units (GPUs) alone can account for about 40%[21]; by 2030, their annual footprint in data centres alone could approach 500 TWh—about 8.6 times Switzerland's current electricity demand[22,23].

However, despite its key role in semiconductor losses, electromagnetic interference (EMI) and other nonidealities that govern the sustainability of electrical and electronic systems[24-26], semiconductor switching has for decades lacked a charge-unified formulation across conceptual, mechanistic, formal and analytical levels. Conceptually, the charge-mediated nature of circuit elements, particularly semiconductor devices, has yet to be elucidated within a unified framework; conventional circuit theory lacks insights on carrier-storage mechanisms and conductive-path evolution in semiconductors and lacks a unified, physics-based definition of turn-on onset (detailed in *Methods*). Mechanistically, unification of macro- and microscopic pictures through switching inertia and interactions is absent; a unified mechanistic account of the physical origins of the spatiotemporal evolution of semiconductor switching remains lacking, with the causal pathways linking input interactions to output responses unresolved, despite substantial efforts to reproduce observed switching waveforms or their linearized approximations.[27-34] For example, the conventional $E_{on}$ model properly accounts for the DC source and output capacitances of both half-bridge devices under iZVS,[35] but not the load current or associated energy dissipation, which leads to prediction errors of 34.41–80.05%. For another example, although the so-called Miller plateau, its abrupt drop (termed Inertia plateau and Inertia drop, respectively within CUSST) and their absence under ZVS have been widely reported[27,28,32-34], their common origin remains obscured by the unrecognized switching inertia and interactions between devices, despite their potentially substantial contributions to system losses, EMI and nonlinearity. In addition, intra-device carrier recovery under ZVS (here termed ZVS recovery) has yet to be revealed, leaving recoveries across switching scenarios unrecognized as different manifestations of the same underlying carrier dynamics and fundamentally ununified. Formally, existing representations remain fragmented. Many of them[27-29] represent the semiconductor device as a two- or three-terminal block; more advanced representations[34,36,37] add junction capacitances to capture dielectric phenomena; some further employ complex, non-unified representations, modelling the conductive path with waveform-fitted current or voltage sources that vary across scenarios and subintervals[25,26,33,38]. Analytically, energy and charge conservation arise from distinct fundamental symmetries: time-translation symmetry through Noether's theorem and electromagnetic gauge symmetry, respectively, with charge conservation expressed in circuit theory through Kirchhoff's current law (KCL). The two frameworks remain separate, and their applicability to switching dynamics has not been established.

Substantial efforts have demonstrated that cross-domain research and co-design can achieve gains beyond individual-domain optimization, including a 50-fold increase in cooling coefficient of performance via electronics-thermal co-design[10], a 15.8-fold enhancement in thermal conductivity via materials–thermal–packaging-application co-design[39], and a twofold increase in energy efficiency via device–circuit–architecture–algorithm research[40]. However, the absence of underpinning theoretical foundations impedes unlocking the full potential of such cross-domain research and co-design along the semiconductor value chain, by obscuring causal links and leaving no unified theoretical language. For example, conventional device-level figure-of-merit (FOM) optimization[41,42] does not reliably improve application-level performance. Although this limitation has been recognized[43,44], mechanistic understanding of switching dynamics remains unresolved, leaving device-to-application causal pathways obscured. As switching heat is highly localized in space and time—within small regions of the die and brief switching subintervals—average junction-temperature estimates may mask much higher local temperature peaks.[45] More broadly, limited understanding of how switching dynamics translate into electrothermal stress and degradation drives reliance on average thermal metrics, empirical derating and conservative margins, leading to material-intensive overdesign while obscuring spatiotemporally localized temperature peaks that may define the true reliability limit.[46-48]

Here we introduce Charge-Unified Semiconductor Switching Theory (CUSST), which establishes a charge-unified formulation across conceptual, mechanistic, formal and analytical levels. Conceptually, CUSST reveals the physical nature of circuit elements, including semiconductor devices, as physical media for charge redistribution and transfer, unifying macro- and microscopic physical pictures. Mechanistically, CUSST reveals switching inertia and its physical origin, switching as a dynamical process, as well as the Inertia plateau, Inertia drop and their absence under ZVS as distinct manifestations of the same underlying dynamics. It further reveals ZVS recovery and the physical unity of carrier mechanisms across switching scenarios. Formally, CUSST-enabled insights are expressed in a representative equivalent-circuit formalism that unifies macro- and microscopic physical pictures. It may serve as a semiconductor counterpart to the standard transformer equivalent-circuit[28,49], in principle generalizable across circuits, devices, scenarios and subintervals. Analytically, CUSST generalizes circuit theory with a charge-unified view, encompassing charge-storage mechanisms and conductive-path resistance in semiconductors. It further unifies charge- and energy-conservation frameworks, establishing their equivalent applicability to switching dynamics. As a result, CUSST enables four fundamental unifications: across charge-storage mechanisms (detailed in Table 3); across charge- and energy-conservation frameworks; across macro- and microscopic insights; and through a unified equivalent-circuit formalism in principle universally applicable across devices, circuits, scenarios and subintervals. These unifications offer unprecedented conceptual, formal and analytical simplicity, immediate practical utility and broader generality, while retaining compatibility with existing theories.

After establishing CUSST, we found that CUSST yielded a new CUSST-based $E_{on}$ model, with a 17-fold prediction accuracy improvement over the conventional model[35] (Fig. 2 and Table 2); CUSST further reveals the causal chain and dynamical mechanism underlying switching-waveform evolution, across scenarios and subintervals (Fig. 3 and Fig. 4), including origins of Miller plateaus and associated gate voltage drops across switching scenarios. Beyond these demonstrated implications, as further discussed in the *Discussion and Outlook*, CUSST formulates the theoretical foundations of a unified system for modelling, design, optimization and standardization across semiconductor science, engineering and downstream applications; CUSST further paves the way for unifying these disparate domains into a unified discipline. For example, by revealing cross-domain causal information, CUSST could enable full-chain causal feedback, cross-domain integrated research and co-design across the semiconductor value chain, unlocking new levels of device- and system-level sustainable performance. A related ZVS study[50] illustrates this potential: by revealing previously unresolved interactions such as opposite-device channel current during turn-off before ZVS—it achieved an approximately 60-fold improvement in average modelling accuracy over prior-art methods, now underpinning ongoing international standardization[51] and advancing the domains across the semiconductor value chain by offering a key piece of causal information.

To our knowledge, CUSST establishes the first unified, global, micro–macro spatiotemporal view of switching throughout the multiscale system and across timescales, moving beyond local, fragmented and domain-isolated descriptions. By offering a unified theoretical language across the semiconductor value chain and cross-domain causal information, CUSST paves the way towards a unified discipline spanning the value chain.

In the following sections, we first present the concept of CUSST (Fig. 1). We then introduce the unified formulation of semiconductor switching dynamics from charge and energy conservation laws, together with experimental validation. Next, we demonstrate the CUSST-enabled causal-mechanistic interpretability by revealing the underlying causal mechanisms and dynamical nature of switching across representative scenarios, including HS and iZVS cases, together with a representative ZVS and an additional iZVS case presented in *Methods*. Finally, we discuss the implications and outlook, followed by *Methods*.

## CUSST concept

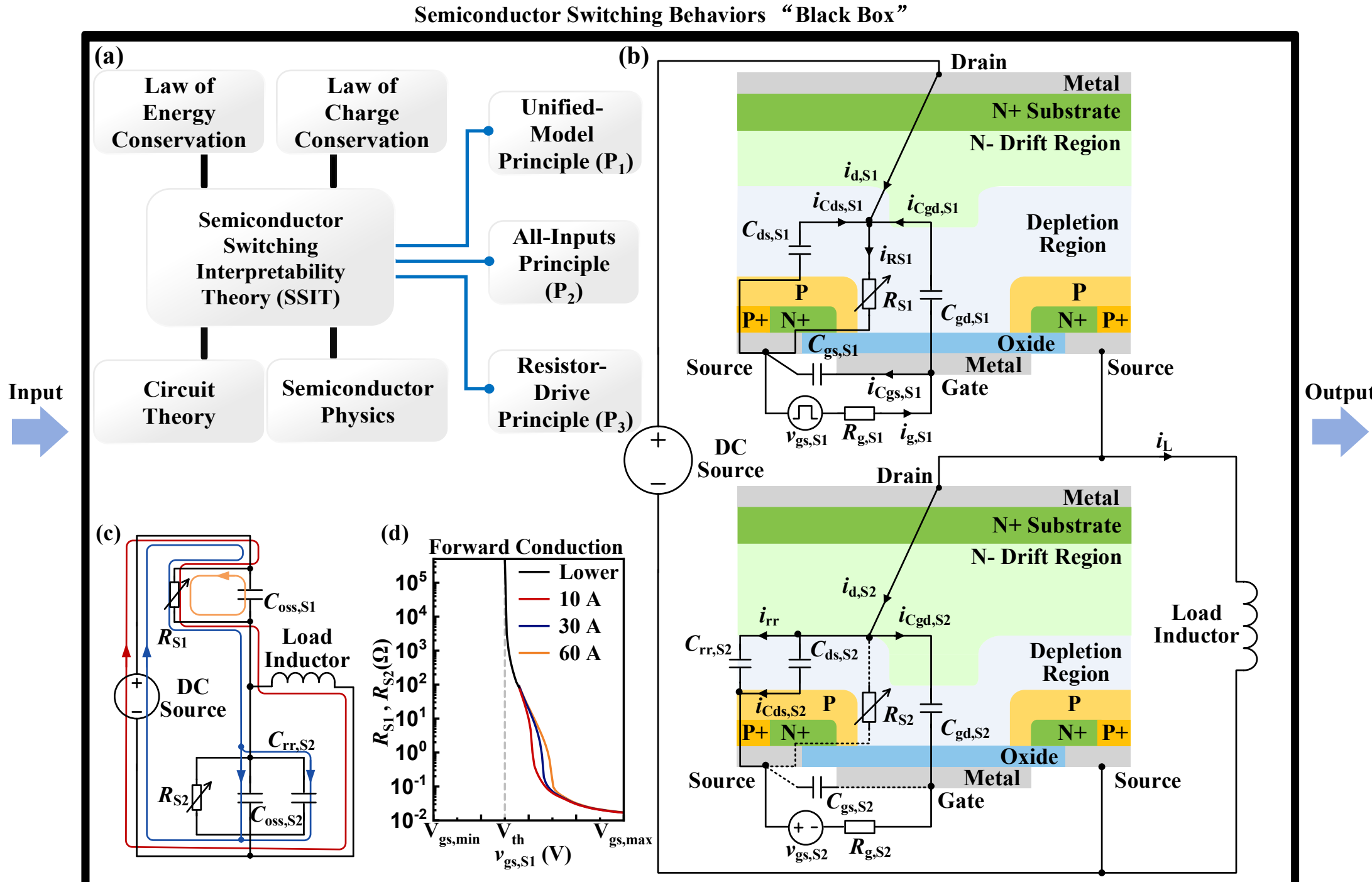

**Fig. 1| Interpretability for semiconductor switching behaviors "black box".** (a) Conceptual Illustration of the role of CUSST to bridge conservation laws of energy and charge, circuit theory and semiconductor physics, together with the three underlying principles of CUSST. (b) Illustration of the unified-model principle using an equivalent-circuit model of a MOSFET half-bridge with a load inductor, showing individual current components at a representative instant during voltage-fall phase (VF) under hard switching. (c) Simplified equivalent-circuit representation of (b), where the red current component denotes the load current; the blue current components denote the charging current component of $S_2$, supplied by the DC source and the orange current component denotes the discharging current of $C_{oss,S1}$ via $S_1$'s channel. (d) $R_{S1}$–$v_{gs,S1}$ curve in forward conduction over the $v_{gs,S1}$ swing, where $V_{gs,max}$ and $V_{gs,min}$ denote the maximum and minimum $v_{gs,S1}$ during the switching transition, respectively. The red, blue and purple curves correspond to conducting current of 10 A, 30 A and 60 A, respectively; the black curve denotes the $R_{S1}$–$v_{gs,S1}$ relationship for $v_{gs,S1}$<4 V at

lower current levels.

CUSST comprises three principles: Switching Inertia Principle, Dynamical Evolution Principle and Interaction Reciprocity Principle. While Fig. 1 shows a MOSFET-based half-bridge with a load inductor as a representative example, the theory is expected to generalize through the underlying switching dynamics, rather than being limited to a specific topology, device type, switching scenario or subinterval.

**1st Principle ($P_1$)—Switching Inertia Principle:** CUSST reveals a charge-unified viewpoint that identifies circuit elements, in their physical nature, as media for charge redistribution and transfer, thereby offering a common conceptual foundation between macroscopic circuit behavior and microscopic semiconductor mechanisms. In particular, semiconductor devices are identified as distributed charge-storing media whose internal charge distributions and dissipative conduction paths undergo a coupled spatiotemporal evolution between insulator-like blocking and conductor-like current-conducting states. Driven by relevant interactions, the underlying charge evolution manifests macroscopically as continuous voltage–current evolution; the equivalent resistance of the conduction path is governed by gate-induced charge and conductivity-modulating stored charge.

As detailed in the *Methods* section on the representative equivalent-circuit formalism, the MOSFET is formalized by a representative equivalent circuit[29,30,52]. This formalism comprises a lumped variable resistor (e.g., $R_{S1}$ and $R_{S2}$), lumped variable capacitors (e.g., $C_{oss,S1}$ and $C_{oss,S2}$) as well as a lumped current source to formalize the conductivity modulation mechanisms. As a semiconductor counterpart to the standard transformers equivalent-circuit formalism[28,49], this representative formalism captures switching dynamics across scenarios and subintervals for the MOSFET and is in principle generalizable across devices and circuits.

During switching, the equivalent resistance evolves between a high-resistance state and a low-resistance state, while charge and energy stored in the capacitive elements and conductivity-modulated regions also evolve with time. At any instant, these resistive, capacitive and carrier-storage states characterize the switching state. Such charge and energy storage gives rise to intrinsic switching inertia: changing capacitive charge storage and, in conductivity-modulated devices, injecting or extracting stored minority carriers require finite charge transfer and associated energy exchange or dissipation, preventing discontinuous state evolution.

**2nd Principle ($P_2$)—Dynamical Evolution Principle:** CUSST reveals the fundamentally dynamical nature of semiconductor switching: stored charges act as momentum-like state variables, while transferred charge acts as an impulse-like quantity. The dimensional correspondence also reflects this shared dynamical nature between mechanics and state evolution of circuit elements.

As demonstrated in Fig.1, the gate current acts as an external gate-driving interaction, analogous to an external force in classical dynamics, while $R_{S1}$ serves as a key switching-state variable. By overcoming switching inertia, this interaction initiates the variation of $R_{S1}$, thereby establishing a causal chain from $R_{S1}$ evolution to charge change, voltage and current evolution, and energy transfer and dissipation. At the macroscopic circuit level, and while incorporating microscopic semiconductor insights, this dynamical evolution is governed by generalized circuit theory within CUSST's charge-unified framework, conservation laws and constrained by the $R_{S1}$ characteristic, nonlinear capacitance characteristics and carrier-storage mechanisms, as demonstrated in the "Causal-Mechanistic Interpretations of Switching Dynamics" sections and relevant sections in *Methods*.

**3rd Principle ($P_3$)—Interaction Reciprocity Principle**: No charge change is one-sided: universally across circuits, devices, scenarios and subintervals, charge changes in one element or charge-storage region are accompanied by coupled responses in other element(s), energy dissipation, or both, under charge and energy conservation. Accordingly, a complete representation of switching dynamics requires all non-negligible interactions to be incorporated; omitting any relevant interaction yields only partial dynamics. In CUSST's charge-unified framework, charge-state changes include inter-element charge transfer and intra-device carrier-storage changes such as recombination. Parasitic inductances are omitted from the formulation, as justified in *Methods*.

**Unified formulation of semiconductor switching dynamics from charge and**

**energy conservation laws and experimental validation**

Within CUSST, an CUSST-based $E_{on}$ model is derived in a representative iZVS scenario, independently from charge and energy conservation (detailed in *Methods*). Considering the same physical interactions, the analytical frameworks based on energy conservation and charge conservation converge to identical analytical expressions, thereby establishing their fundamental equivalence in semiconductor switching.

We further validate the CUSST-based model against experimental measurements and compare it with the conventional model[35]. The overall data comparison is visualized in

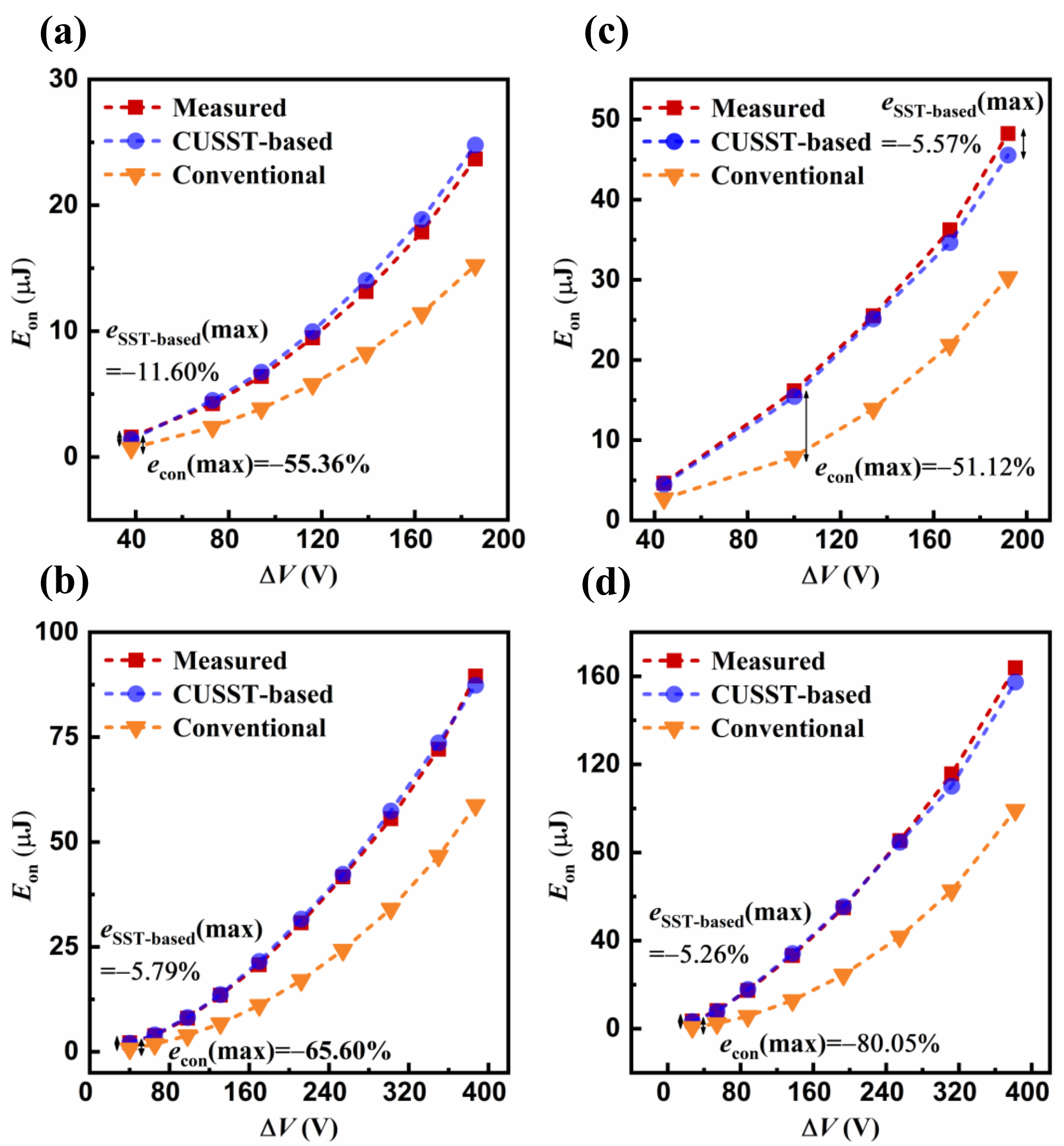

Fig. **2**, while the error-reduction factor is quantified for each condition in

Table 2. Averaging these error-reduction factors across all conditions gives a 17-fold average error reduction relative to the conventional model[35]. This agreement with experimental measurements validates the equivalent applicability of the two analytical frameworks to semiconductor switching and supports the deterministic predictive capability of CUSST. These results further confirm CUSST's compatibility with charge and energy conservation laws, extending their applicability to semiconductor devices.

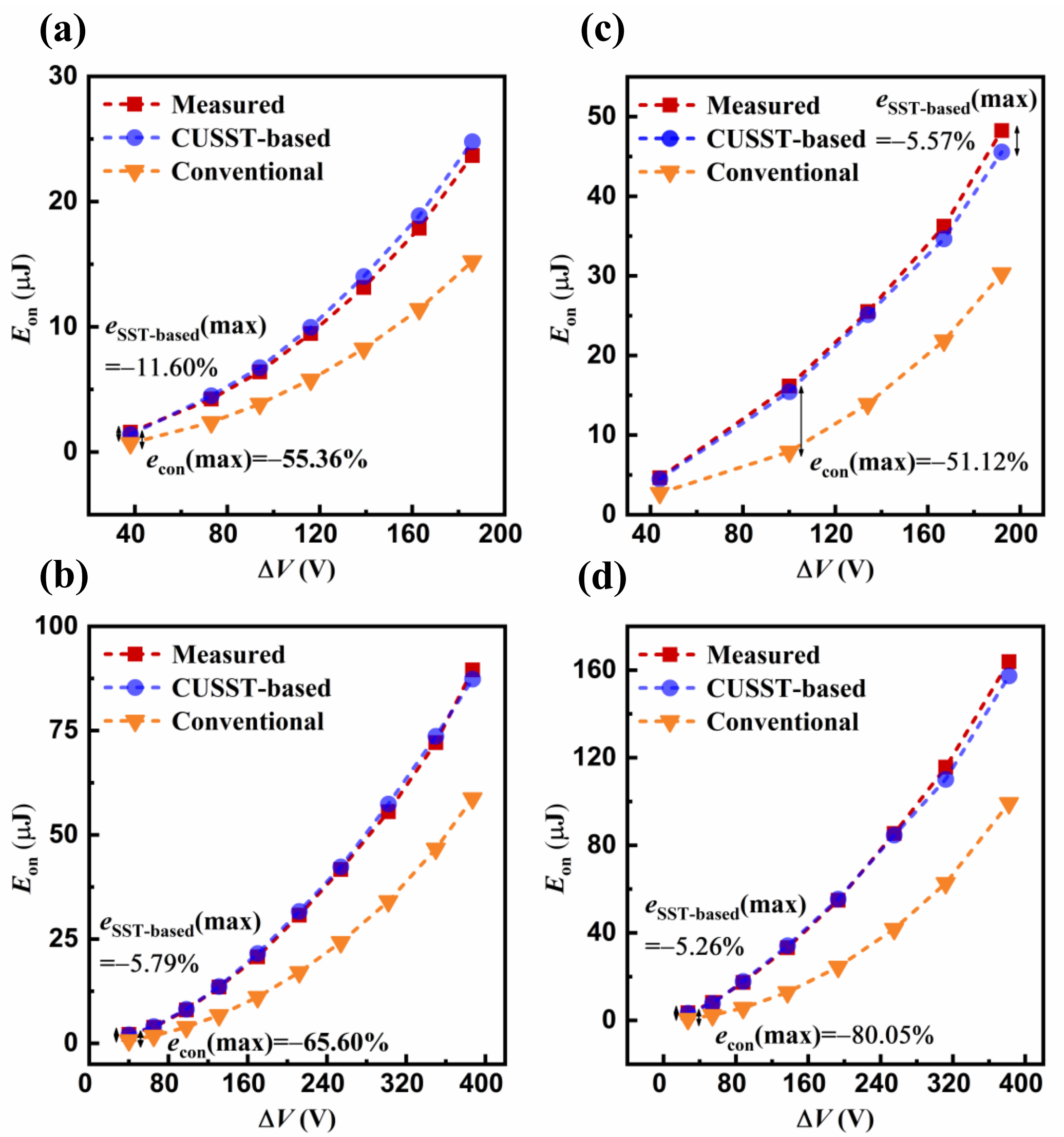

**Fig. 2| Comparison of measured results (red) with calculated results from the CUSST-based model (blue) and the conventional model (orange) at room temperature.** Measurements were performed using commercially available semiconductor devices. Results are shown for (a) $V_{DC}$=200 V with semiconductor device A[53]; (b) $V_{DC}$=400 V with semiconductor device A[53]; (c) $V_{DC}$=200 V with semiconductor device B[54] and (d) $V_{DC}$=400 V with semiconductor device B[54].

## Revealing causal mechanisms and the dynamical nature of representative HS through CUSST-enabled causal-mechanistic interpretability

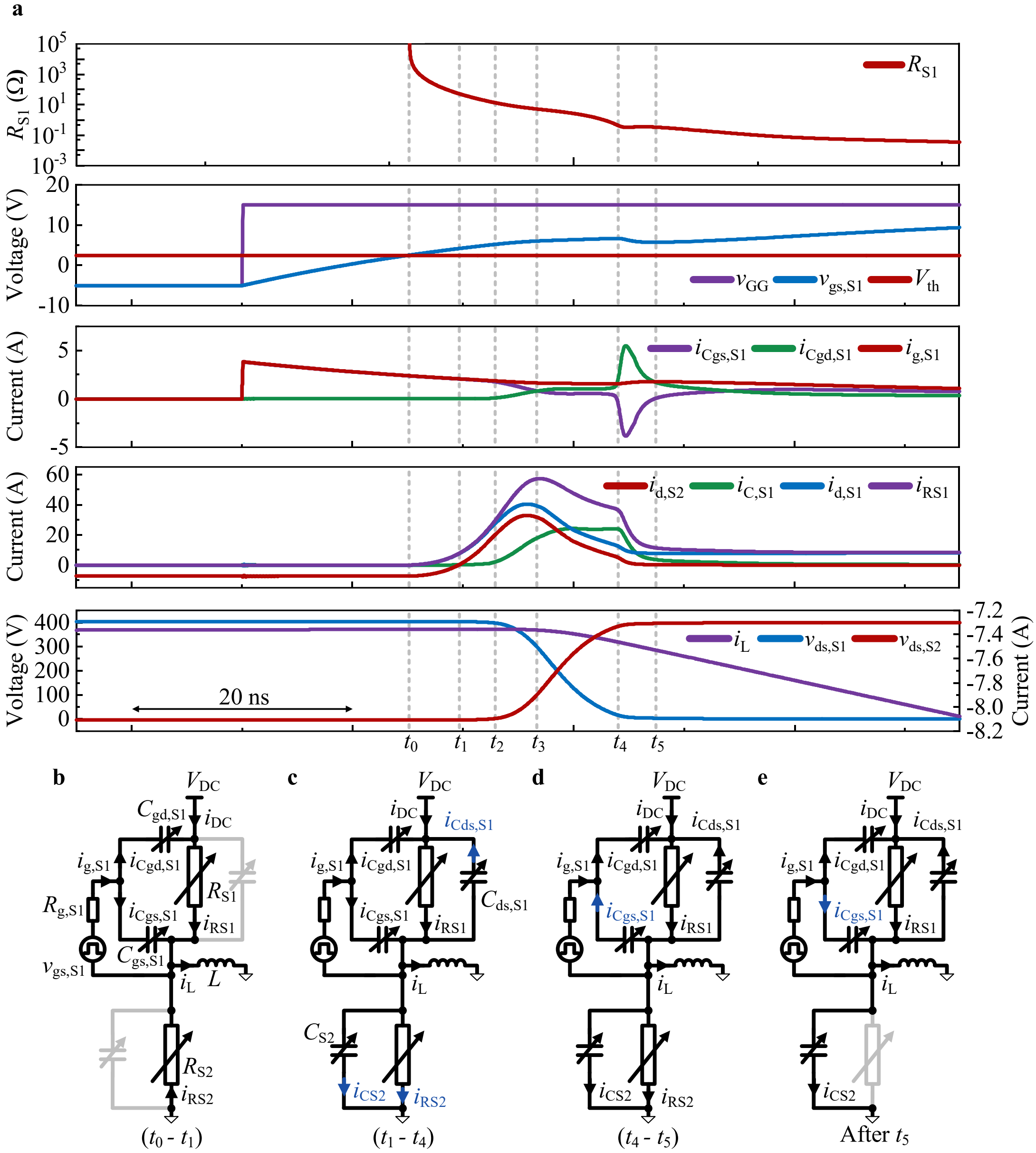

**Fig. 3 | Revealing causal mechanisms and the dynamical nature of representative hard switching through CUSST-enabled causal-mechanistic interpretability a,** switching waveforms within a representative HS. **b,** Operation mode during subinterval ($t_0$-$t_1$). **c,** Operation mode during subinterval ($t_1$-$t_4$). **d,** Operation mode during subinterval ($t_4$-$t_5$). **e,** Operation mode after $t_5$. The equivalent-circuit formalism is obtained by $\mathbf{P_1}$; the highlighted branches in **b-e** indicate the current paths; the current components that differ from those in the immediately preceding sub-figure are highlighted in blue.

At $t_0$ (i.e., the onset of HS), $i_L$ is reverse-conducted by $R_{S2}$. During ($t_0$ - $t_1$), while $v_{ds,S1}$ remains nearly constant, $v_{gs,S1}$ increases and $R_{S1}$ therefore decreases rapidly accordingly; hence, $i_{RS1}$ increases (by $\mathbf{P_2}$ and Ohm's law). As a result, a lossy current-commutation phase (CC) occurs, in which $i_L$ commutates from $R_{S2}$ to $R_{S1}$ (by $\mathbf{P_3}$ and KCL; with charge conveyed by $i_L$ considered).

At $t_1$, $i_L$ is entirely conducted by $R_{S1}$, which indicates the CC's completion; a reverse-recovery process (RR) and a VF commence simultaneously. After $t_1$, during RR, the stored minority carriers in $S_2$ are removed by the combined effects of recombination and $i_{RS2}$ whose direction is reversed at $t_1$; meanwhile, during VF, $i_{CS2}$ charges $C_{S2}$ via $R_{S1}$, which increases $v_{ds,S2}$. Since $V_{DC}$ remains constant and $V_{DC}=v_{ds,S1}+v_{ds,S2}$, $v_{ds,S1}$ drops accordingly; therefore, $C_{S1}$ discharges via $R_{S1}$ (by $\mathbf{P}_3$, considering influence of DC source). Notably, during ($t_1$-$t_5$), $|dv_{ds,S2}/dt|=|i_{d,S2}-i_{RS2}|/C_{S2}$.

During ($t_1$-$t_3$), a quicker relative drop in $R_{S1}$ compared to $v_{ds,S1}$ (i.e., $|dR_{S1}/R_{S1}|>|dv_{ds,S1}/v_{ds,S1}|$) causes a continued increase in $i_{RS1}$, which in turn increases $i_{d,S2}$ (by $\mathbf{P}_3$ and Ohm's law); meanwhile, as $v_{gs,S1}$ increases, $|dR_{S1}/R_{S1}|$ decreases, which further leads to a reduced *di/dt* of $i_{RS1}$ (by $\mathbf{P}_2$; considering $R_{S1}$'s nonlinear characteristic).

During ($t_1$-$t_2$), by $\mathbf{P}_3$ and with the nonlinearity of $C_{oss,S2}$ considered, $S_2$'s *dv/dt* remains low because $C_{oss,S2}$ is still large, although $C_{oss,S2}$ decreases as $v_{ds,S2}$ increases and $i_{d,S2}$ rises. Therefore, only a small portion of $i_{g,S1}$ is required by $C_{gd,S1}$ to follow $S_2$'s *dv/dt*; consequently, most of $i_{g,S1}$ charges $C_{gs,S1}$.

During ($t_2$-$t_3$), $C_{oss,S2}$ transitions from its high- to low-capacitance region, which leads to an increase in $S_2$'s *dv/dt* (by $\mathbf{P}_3$; with $C_{oss,S2}$'s nonlinearity considered). As $C_{oss,S1}$'s *dv/dt* dynamically follows $S_2$'s *dv/dt*, while $C_{ds,S1}$ increases as $v_{ds,S1}$ decreases, $i_{Cds,S1}$ rises as a result (by Kirchhoff's voltage law (KVL); with $C_{oss,S1}$'s nonlinearity considered). A feedback mechanism occurs to overcome the switching inertia, owing to a large amount of charge required by $C_{gd,S1}$, more $i_{g,S1}$ is diverted to $C_{gd,S1}$, which leads to: (1) higher $i_{Cgd,S1}$, which increases $C_{gd,S1}$'s *dv/dt*; and (2) lower $i_{Cgs,S1}$, which slows the rise of $v_{gs,S1}$, the reduction of $R_{S1}$, and consequently the rise of $i_{RS1}$ (by $\mathbf{P}_2$ and Ohm's law). The slower rise of $i_{RS1}$, together with the significant rise of $i_{C,S1}$, reduces the *di/dt* of $i_{d,S2}$, which is initially positive but becomes negative before $t_3$. This limits the rise of $S_2$'s *dv/dt* despite the decrease in $C_{oss,S2}$(by $\mathbf{P}_3$; with $C_{oss,S2}$'s nonlinearity considered). Therefore, because of this feedback mechanism, $C_{gd,S1}$'s *dv/dt* dynamically follows $S_2$'s *dv/dt*.

During ($t_3$-$t_4$), a slower relative drop in $R_{S1}$ compared to $v_{ds,S1}$ (i.e., $|dR_{S1}/R_{S1}|<|dv_{ds,S1}/v_{ds,S1}|$), causes a decrease in $i_{RS1}$ (by $\mathbf{P}_2$ and Ohm's law). Initially,

$C_{ds,S1}$ and $C_{gd,S1}$ increase (by $\mathbf{P}_3$; with their nonlinearity considered), whereas $S_1$'s *dv/dt* changes only slightly; thus $i_{C,S1}$ rises briefly; and slightly more $i_{g,S1}$ is diverted to $C_{gd,S1}$. Initially, the fall in $i_{RS1}$ and the rise in $i_{C,S1}$ both reduces $i_{d,S2}$. Thereafter, the continued fall in $i_{RS1}$ further reduces $i_{d,S2}$ and lowers $S_2$'s *dv/dt*, due to a quicker relative drop in $i_{d,S2}$ compared to $C_{oss,S2}$ (i.e., $|di_{d,S2}/i_{d,S2}|>|dC_{oss,S2}/C_{oss,S2}|$). Because $S_1$'s *dv/dt* follows $S_2$'s *dv/dt*, $S_1$'s *dv/dt* also decreases; while the *dv/dt* of $C_{ds,S1}$ and $C_{gd,S1}$ decreases, $C_{ds,S1}$ and $C_{gd,S1}$ increase with the decreased $v_{ds,S1}$; thus $i_{Cds,S1}$ and $i_{Cgd,S1}$ nearly stabilize (by $\mathbf{P}_3$; with nonlinearity of $C_{oss,S1}$ and $C_{oss,S2}$ considered).

During ($t_4$-$t_5$), as $v_{ds,S1}$ falls, $C_{gd,S1}$ enters its high-capacitance region and rises sharply; meanwhile, as $v_{ds,S2}$ approaches $V_{DC}$, $C_{oss,S2}$ remains low, and $i_{d,S2}$ sustains a significant $S_2$'s *dv/dt*. Therefore, $i_{Cgd,S1}$ must remain sufficient to sustain a comparable $C_{gd,S1}$ *dv/dt* (by $\mathbf{P}_3$ and KVL; with the nonlinearity of $C_{gd,S1}$ and $C_{oss,S2}$ considered). Because $i_{g,S1}$ is limited, a stronger feedback mechanism occurs and causes $v_{gs,S1}$ to drop. This $v_{gs,S1}$'s drop raises $R_{S1}$ and boosts $i_{g,S1}$; together with the reduced $v_{ds,S1}$, the increased $R_{S1}$ lowers $i_{RS1}$ (by $\mathbf{P}_2$ and Ohm's law), suppresses $i_{d,S2}$ and thus limits $S_2$'s *dv/dt*. Meanwhile, a supplementary charge impulse is extracted from $C_{gs,S1}$ to help to discharge $C_{gd,S1}$ via $R_{S1}$ (by $\mathbf{P}_3$). Together, these effects supply the required charge impulse through $i_{Cgd,S1}$ to $C_{gd,S1}$ while they restrain $S_2$'s *dv/dt*, such that $C_{gd,S1}$'s *dv/dt* follows $S_2$'s *dv/dt*. This feedback mechanism also reveals the origin of the Miller plateau drop.

After $t_5$, as $v_{gs,S1}$ further increases, $R_{S1}$ decreases slowly (by $\mathbf{P}_2$), which causes a slight drop in the half-bridge midpoint voltage and thus a minor discharge of $C_{S1}$ and charge of $C_{S2}$ via $R_{S1}$, respectively.

**Revealing causal mechanisms and the dynamical nature of representative iZVS through CUSST-enabled causal-mechanistic interpretability**

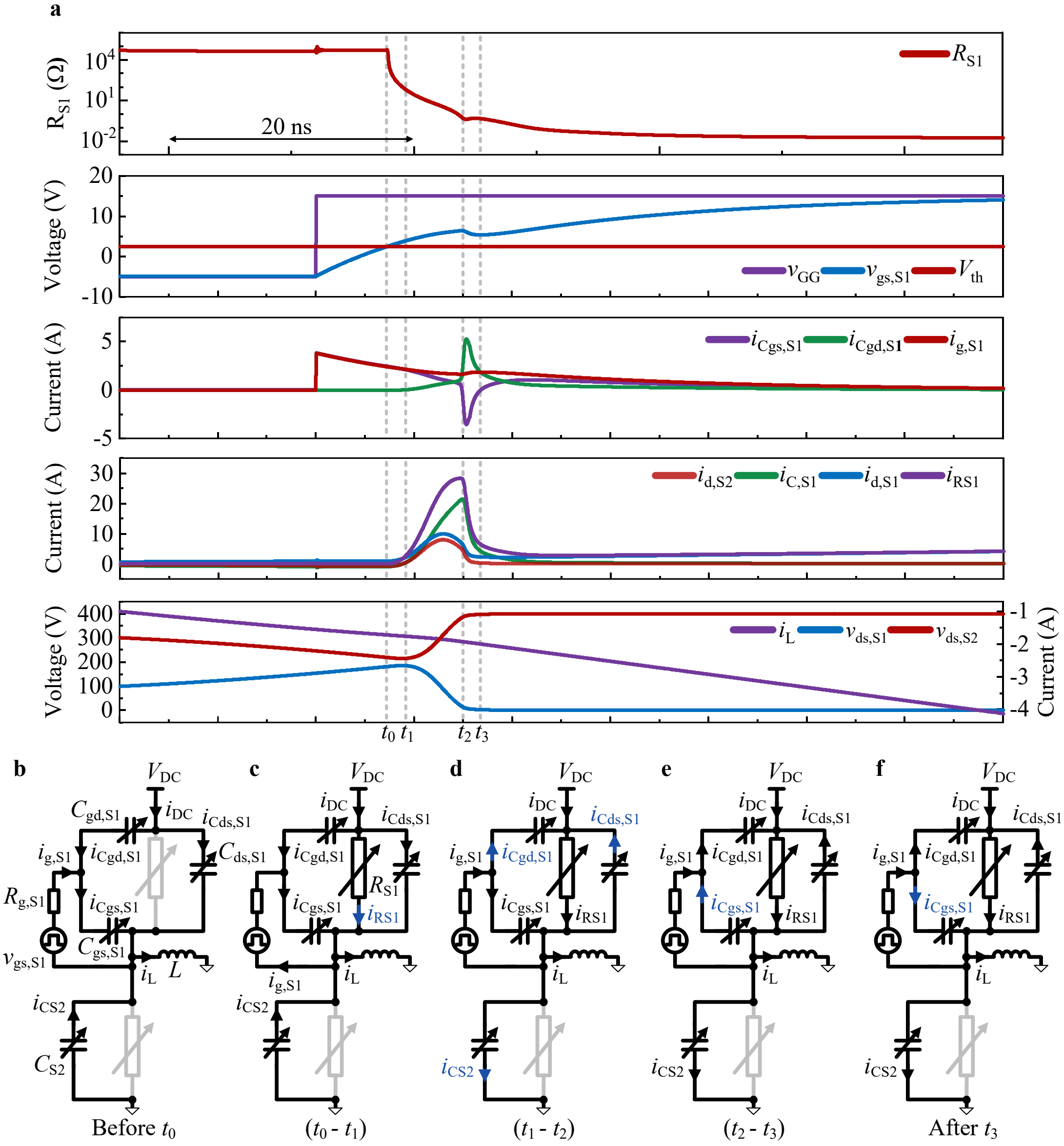

**Fig. 4| Causal-Mechanistic Interpretability of Switching Dynamics in a Representative iZVS Scenario. a**, Switching waveforms in a representative iZVS scenario ($i_L$ flows out from the half-bridge midpoint throughout the entire iZVS). **b**, Operation mode before $t_0$. **c**, Operation mode during subinterval ($t_0$-$t_1$). **d**, Operation mode during subinterval ($t_1$-$t_2$). (e) Operation mode ($t_2$-$t_3$). (f) Operation mode after $t_3$. The equivalent-circuit formalism is obtained by **P₁;** the highlighted branches in **b-f** indicate the current paths; the current components that differ from those in the immediately preceding sub-figure are highlighted in blue.

Before $t_0$ (i.e., the onset of the iZVS), $i_{\mathrm{L}}$ charges $C_{\mathrm{S1}}$ and discharges $C_{\mathrm{S2}}$ simultaneously. During ($t_0$-$t_1$), $i_{\mathrm{L}}$ continues to charge $C_{\mathrm{S1}}$ and discharge $C_{\mathrm{S2}}$ simultaneously; as $v_{\mathrm{gs,S1}}$ has exceeded $V_{\mathrm{th,S1}}$ and continues to increase, $R_{\mathrm{S1}}$ decreases rapidly, which increases $i_{\mathrm{RS1}}$ (by $\mathbf{P_2}$ and Ohm's law). Consequently, $i_{\mathrm{L}}$ gradually commutates to $R_{\mathrm{S1}}$, till $t_1$, when $i_{\mathrm{L}}$ is fully conducted by $R_{\mathrm{S1}}$ (by $\mathbf{P_3}$ and KCL; with charge conveyed by $i_{\mathrm{L}}$ considered).

During ($t_1$-$t_2$), initially, as $v_{gs,S1}$ increases, $R_{S1}$ decreases, which increases $i_{RS1}$ (by **P2** and Ohm's law). As a result, $i_{CS2}$ increases; together with a decrease of $C_{oss,S2}$, it leads to an increase in $S_2$'s *dv/dt* (by **P2**; with $C_{oss,S2}$'s nonlinearity considered). A feedback mechanism occurs to overcome $S_1$'s switching inertia, as $C_{gd,S1}$ requires a large charge rate to follow the *dv/dt* evolution (by **P1**, **P2** and KVL). In this feedback, more $i_{g,S1}$ diverts to $C_{gd,S1}$, which increases $i_{Cgd,S1}$, and thus promotes $C_{gd,S1}$'s *dv/dt*; meanwhile, $i_{Cgs,S1}$ decreases, which slows the rise of $v_{gs,S1}$, the reduction of $R_{S1}$, and thus the increase of $i_{RS1}$ (by **P2** and Ohm's law). Slower $i_{RS1}$ rise, together with a rapid rise of $i_{CS1}$, suppresses *di/dt* of $i_{d,S2}$, which is initially positive, but becomes negative before $t_2$, and thus limits $S_2$'s *dv/dt* despite the decrease of $C_{oss,S2}$ (by **P2** and KCL; with $C_{oss,S2}$'s nonlinearity considered). This feedback satisfies $C_{gd,S1}$'s charge-rate requirement to follow $S_2$'s *dv/dt* evolution.

During ($t_2$-$t_3$), $C_{oss,S2}$ remains low as $v_{ds,S2}$ approaches $V_{DC}$; $S_2$'s *dv/dt* remains significant despite limited $i_{d,S2}$. To follow $S_2$'s *dv/dt*, $C_{gd,S1}$ requires a more significant charge rate as $C_{gd,S1}$ enters its high-capacitance region, and continues to increase as $v_{ds,S1}$ decreases (by **P2** and KVL; with $C_{gd,S1}$'s and $C_{oss,S2}$'s nonlinearity considered). As $i_{g,S1}$ alone cannot satisfy $C_{gd,S1}$'s higher charge-rate requirement, a further feedback mechanism occurs to overcome $S_1$'s greater inertia (by **P1** and **P2**). In this feedback, $C_{gs,S1}$ transfers a charge impulse to discharge $C_{gd,S1}$ via $S_1$'s channel, which causes a drop in $v_{gs,S1}$. This raises $i_{g,S1}$, increases $i_{Cgd,S1}$, and increases $R_{S1}$; together with the fall of $v_{ds,S1}$, $i_{RS1}$ rapidly decreases (by **P2** and Ohm's law). Consequently, $i_{d,S2}$ is suppressed, which directly limits $S_2$'s *dv/dt*. This feedback satisfies $C_{gd,S1}$'s charge-rate requirement to follow $S_2$'s *dv/dt* evolution, and also reveals the origin of the Miller plateau drop.

After $t_3$, $C_{gs,S1}$ no longer discharges; instead, $i_{g,S1}$ charges both $C_{gs,S1}$ and $C_{gd,S1}$ simultaneously. As $v_{gs,S1}$ further increases, $R_{S1}$ decreases slowly. Despite low $R_{S1}$, $i_{RS1}$ remains low as $v_{ds,S1}$ is already low (by **P2** and Ohm's law). Excluding the discharging current of $C_{oss,S1}$, $i_{d,S2}$ is also low, which causes only a slow and slight rise in the half-bridge midpoint voltage, accompanied by minor discharge of $C_{S1}$ and charge of $C_{S2}$ via $R_{S1}$.

**Discussion and outlook**

We present CUSST as a general theory at conceptual, mechanistic, formal and analytical levels. CUSST reveals a charge-unified physical view of circuit elements, particularly semiconductor devices, as media for charge redistribution and transfer, thereby establishing a unified physical picture of switching across microscopic and macroscopic scales and throughout its spatial and temporal evolution. CUSST further reveals switching inertia and its origin, and identifies the nature of switching as a dynamical process in which physical interactions overcome this inertia. To formalize CUSST-enabled insights, we present a representative physics-based equivalent-circuit formalism. CUSST further links switching dynamics to classical mechanics, suggesting that their dimensional correspondence is a manifestation of the underlying dynamical nature shared across disciplines. Therefore, CUSST establishes a unified, global, micro–macro spatiotemporal view of switching, moving beyond local and fragmented descriptions, while bridging circuit theory[52], conservation laws of energy and charge, and semiconductor physics. As a result, CUSST enables four fundamental unifications: in the nature of circuit elements via a charge-unified viewpoint; between charge- and energy-conservation frameworks; across macroscopic circuit-level and microscopic semiconductor-physics insights; and through three principles formulated to be generalizable across circuits, devices, scenarios and subintervals, providing a unifying basis under which equivalent-circuit formalisms can be constructed, with their applicability to other contexts open for future exploration. These unifications help to generalize circuit theory to all circuit elements, including semiconductor devices, while conferring unprecedented simplicity at the conceptual, formal and analytical levels and maintaining compatibility with existing theories and practical usability.

As demonstrated implications, we found that CUSST reveals the causal chain and dynamical mechanism underlying switching-waveform evolution across representative switching scenarios, with macro- and microscopic physical pictures unified and significant interactions accounted for. We further found that CUSST offers deterministic predictive capability through an CUSST-based $E_{\mathrm{on}}$ model, which achieves a 17-fold average error reduction compared to the conventional model[35]. In addition, we validated the fundamental applicability and equivalence of the energy- and charge-conservation frameworks in switching dynamics, revealing complementary perspectives on switching dynamics: energy conservation captures net energy redistribution and dissipation, while charge conservation highlights the spatiotemporal charge evolution. These results confirm CUSST's compatibility with conservation laws, extending them to semiconductor devices. Together, these representative examples suggest that CUSST provides a basis for a new theoretical system.

Beyond the demonstrated examples, CUSST may have broader and longer-term implications whose full scope remains difficult to assess at this stage. For example, CUSST may help to identify directions across disciplines such as semiconductor materials[4], chip design[5], packaging[6], reliability[7,8], thermal engineering[9,10]; in applications such as power electronics; and potentially across broader systems, including communications, computation devices and integrated circuits.[11,12] More broadly, CUSST could support the design, optimization and standardization across semiconductor science, engineering and downstream applications, as illustrated by the related ZVS study[50], which improved modelling accuracy 60-fold, now underpins ongoing international standardization and advances domains across the semiconductor value chain. Beyond this, CUSST could help unify these domains into a unified discipline across the semiconductor value chain including through enabling joint study, co-design, co-optimization, future full-chain causal feedback mechanisms and cross-domain research. For example, CUSST may translate application-level causal information such as constraints on loss, electrothermal stress and reliability into actionable priorities for upstream semiconductor materials, chip design, fabrication, packaging and reliability engineering, as well as downstream topology, layout, gate-drive, control and thermal co-design. CUSST-enabled, causally informed full-chain reliability co-design—guided by spatiotemporal thermal evolution and highly localized temperature peaks—could right-size die and cooling requirements, reducing overdesign while improving efficiency, environmental performance, lifespan and economic viability. As a specific case, CUSST may shift semiconductor optimization beyond conventional FOM optimization towards customized design weights. For example, in ZVS-capable applications, CUSST-informed weighting of residual switching and conduction losses may prioritize lower drift-region resistance in materials and chip design, alongside layouts that minimize parasitic capacitance. Besides, CUSST could identify underlying causal mechanisms in loss- or electrothermal-stress-dominant switching subintervals and translate them into upstream materials, chip-design and packaging targets, while enabling reliability engineering to address associated electric-field, thermal-cycling and interface-degradation mechanisms beyond conventional average thermal metrics, thereby improving lifetime and cost-effectiveness. By enabling cross-domain causal links across the semiconductor value chain, CUSST could contribute to broader sustainability gains, as illustrated by prior achievements including a 50-fold increase in cooling coefficient of performance via electronics-thermal co-design[10], a 15.8-fold increase in thermal conductivity via materials–thermal–packaging-application co-design[39], and a twofold increase in energy efficiency via device–circuit–architecture–algorithm research[40], suggesting substantial opportunities across CUSST-enabled interfaces. We anticipate that CUSST, along with future developments, could extend to broader applications, including computing, communications and integrated circuits, supporting sustainable performance in systems key to AI, digitalization and electrification.

### Extended Figure - Experimental platform

**Extended Fig. 1.** Experimental platform for $E_{on}$ measurement, using a power electronic converter as the demonstrative example. **(a)** Overview picture of the entire experimental platform. **(b)** Close-up picture of the power electronic converter.

Table 2| Detailed data comparison of $E_{on}$ measured results with calculated values using the CUSST-based model and conventional model, respectively

| Device conduction resistance (mΩ) | $V_{DC}$ (V) | $\Delta V$ (V) | Measured results (μJ) | Calculated values using conventional model (μJ) with error ($e_{\text{con}}$) | Calculated values using CUSST-based model (μJ) with error ($e_{\text{CUSST-based}}$) | Error reduction $\lvert e_{\text{con}}/e_{\text{CUSST-based}}\rvert$ |
|---|---|---|---|---|---|---|
| 25 | 200 | 44 | 4.65 | 2.74 (–41.11%) | 4.49 (–3.57%) | 11.50 |
| 25 | 200 | 100 | 16.17 | 7.91 (–51.12%) | 15.45 (–4.50%) | 11.35 |
| 25 | 200 | 134 | 25.53 | 13.91 (–45.52%) | 25.14 (–1.53%) | 29.67 |
| 25 | 200 | 167 | 36.25 | 21.88 (–39.64%) | 34.65 (–4.41%) | 8.98 |
| 25 | 200 | 192 | 48.25 | 30.31 (–37.19%) | 45.57 (–5.57%) | 6.68 |
| 25 | 400 | 27 | 3.52 | 0.70 (–80.05%) | 3.33 (–5.26%) | 15.21 |
| 25 | 400 | 54 | 8.25 | 2.39 (–71.02%) | 7.87 (–4.58%) | 15.49 |
| 25 | 400 | 88 | 17.36 | 5.75 (–66.91%) | 17.88 (2.95%) | 22.66 |
| 25 | 400 | 137 | 33.21 | 12.91 (–61.11%) | 34.14 (2.82%) | 21.67 |
| 25 | 400 | 193 | 54.94 | 24.46 (–55.48%) | 55.46 (0.94%) | 58.75 |
| 25 | 400 | 255 | 85.42 | 41.79 (–51.08%) | 84.67 (–0.88%) | 57.75 |
| 25 | 400 | 312 | 115.68 | 62.76 (–45.75%) | 110.10 (–4.82%) | 9.49 |
| 25 | 400 | 382 | 163.86 | 99.24 (–39.44%) | 157.42 (–3.93%) | 10.02 |
| 80 | 200 | 38 | 1.59 | 0.709 (–55.36%) | 1.40 (–11.60%) | 4.77 |
| 80 | 200 | 73 | 4.24 | 2.38 (–43.78%) | 4.50 (6.10%) | 7.17 |
| 80 | 200 | 94 | 6.40 | 3.85 (–39.77%) | 6.74 (5.40%) | 7.37 |
| 80 | 200 | 116 | 9.47 | 5.78 (–38.91%) | 9.96 (5.18%) | 7.51 |
| 80 | 200 | 139 | 13.14 | 8.26 (–37.13%) | 14.02 (6.70%) | 5.54 |
| 80 | 200 | 163 | 17.88 | 11.42 (–36.11%) | 18.87 (5.56%) | 6.49 |
| 80 | 200 | 186 | 23.68 | 15.23 (–35.69%) | 24.79 (4.69%) | 7.61 |
| 80 | 400 | 40 | 2.17 | 0.748 (–65.60%) | 2.05 (–5.79%) | 11.33 |
| 80 | 400 | 65 | 3.89 | 1.82 (–53.14%) | 4.06 (4.50%) | 11.81 |
| 80 | 400 | 98 | 8.03 | 3.92 (–51.21%) | 8.20 (2.12%) | 24.13 |
| 80 | 400 | 131 | 13.50 | 6.77 (–49.82%) | 13.66 (1.25%) | 39.99 |
| 80 | 400 | 170 | 20.77 | 11.13 (–46.40%) | 21.54 (3.73%) | 12.45 |
| 80 | 400 | 212 | 30.73 | 17.05 (–44.53%) | 31.65 (2.98%) | 14.93 |
| 80 | 400 | 254 | 41.71 | 24.28 (–41.79%) | 42.34 (1.51%) | 27.70 |
| 80 | 400 | 302 | 55.54 | 34.10 (–38.61%) | 57.40 (3.34%) | 11.54 |
| 80 | 400 | 350 | 72.14 | 46.67 (–35.31%) | 73.67 (2.11%) | 16.70 |
| 80 | 400 | 387 | 89.57 | 58.75 (–34.41%) | 87.35 (–2.49%) | 13.85 |

Table 3| CUSST's charge-unified framework for semiconductor switching dynamics

| Dynamical role | Charge Unified framework | Linear or nonlinear capacitive storage | Carrier-storage region |
|---|---|---|---|
| Momentum-like state variable | Stored charge $Q$ as general state variable | $Q_x=\int C_x dv$ | Stored carrier charge |
| Force-like state rate | $i_x=dQ_x/dt$ | $i_x=dQ_x/dt= C_x(v)dv/dt$ | $dQ_x/dt=i_{inj}- i_{ext}-i_{rec}$ |
| Impulse-like state change | $\Delta Q_x=\int i_x dt$ | $\Delta Q_x=\int C_x dv$ | $\Delta Q_x=\int(i_{inj}-i_{ext}-i_{rec})dt$ |
| Underlying mechanism | Charge transfer or recombination | Charge transfer changes capacitive voltage | Injected, extracted or recombined charge changes |
| Energy association | Energy exchange or | $E_x=\int v dQ_x=\int v C_x(v)dv$ | Energy exchange or |

| | dissipation | | dissipation |
|---|---|---|---|
| Associated state variables | All state variables | Variable Capacitances | Variable resistance (combined with gate-induced charge storage) |
| Role in CUSST | Unified and more fundamental framework | Generalized capacitive case | Generalized carrier-storage case |

# Methods

## A representative equivalent-circuit formalism for switching dynamics

Taking a MOSFET-based half-bridge as an example, the equivalent-circuit formalism of the conductive current path of $S_1$ is represented by a single lumped variable resistor,[29,30,52] whose resistance is defined by Ohm's law as $R_{S1}=v_{ds,S1}/i_{RS1}$, where $i_{RS1}$ denotes the conductive current component through the lumped resistive path. The drain current also includes capacitive current components associated with the capacitive elements.

$R_{S1}$ represents the lumped resistance of the conductive path(s), aggregating relevant resistive elements, including the channel resistance associated with the gate-induced charge storage, accumulation-region resistance, JFET-region resistance, drift-region resistance and other resistive components, some of which may be affected by conductivity-modulation carrier storage in bipolar devices, together with contact resistance.[27] In this work, the $R_{S1}$ formalism is demonstrated for MOSFETs; its extension to other device types, conduction modes and gate-bias conditions is expected in principle but requires future validation. Its I–V characteristic follows the static output characteristic in forward conduction and the static third-quadrant characteristic in reverse conduction, at the corresponding gate voltage. At any operating point, $R_{S1}$ is given by the reciprocal slope of the line connecting the operating point and the origin of the I-V curve at the corresponding gate voltage. The same formalism applies to $R_{S2}$. For example, during a half-bridge shoot-through event, an additional shoot-through current component flows through both conductive paths, represented by $R_{S1}$ and $R_{S2}$.

Regarding capacitive current paths, variable-capacitance equivalent-circuit representations are used.[29,30,52] For example, in a MOSFET-based half-bridge, the displacement current components associated with $C_{ds,S1}$ and $C_{gd,S1}$ traverse the channel, accumulation region, JFET region and drift region. Since $R_{S1}$ primarily represents the aggregate resistance of these regions, both $C_{ds,S1}$ and $C_{gd,S1}$ can be approximated, at the lumped-circuit level, as capacitive paths directly in parallel with $R_{S1}$. For simplicity in this representative formalism, they are combined into a single lumped output capacitance $C_{oss,S1}$, directly in parallel with $R_{S1}$.

Whereas channel conduction in a unipolar device such as the MOSFET does not involve minority-carrier injection or conductivity modulation, conduction through its intrinsic body diode can involve minority-carrier injection and consequent conductivity modulation. Accordingly, during HS of $S_1$, $S_1$ does not intrinsically contain stored minority-carrier charge, whereas the body diode of $S_2$ may contain stored charge if it has previously conducted. The subsequent reverse bias of $S_2$ then induces reverse recovery, producing a carrier-extraction current through the conductive path represented by $R_{S2}$, in addition to the displacement current through $C_{oss,S2}$. A current source in parallel with $R_{S2}$ is therefore enabled to represent this reverse-recovery current component in this case.

During ZVS of $S_1$, by contrast, the body diode of $S_1$ may have conducted before channel turn-on and may therefore contain stored minority-carrier charge. As the diode current decreases, this stored charge must be removed. The associated carrier-removal current, here termed the ZVS-recovery current, flows through the conductive path represented by $R_{S1}$ and is formalized using a current source in parallel with $R_{S1}$. Here, ZVS recovery denotes carrier removal during the ZVS commutation process rather than conventional reverse recovery under substantial reverse bias in the HS process.

In principle, the forward conduction of bipolar or conductivity-modulated devices such as IGBTs can also contribute to the conductive current component, while the validation for such devices remains future work.

**Omission of parasitic inductances from the CUSST formulation**

Parasitic inductances inevitably exist in semiconductor devices, layouts and other circuit elements, and their inclusion would provide a closer representation of practical conditions. Their omission from the CUSST formulation is analogous to the idealized formulation of inertial motion in classical mechanics, where resistive effects such as air resistance exist in practice but are omitted to reveal the intrinsic dynamics of motion. This treatment isolates intrinsic switching dynamics and avoids obscuring the underlying switching causality. Furthermore, the parasitic inductances associated with packages and layouts are typically orders of magnitude larger than the parasitic inductance associated with the semiconductor device itself, and therefore usually dominate the overall parasitic inductance. Significant advances in packaging and layout technologies have nevertheless pushed the overall parasitic inductance towards increasingly low levels, with reported values already below 1 nH in recent literature[6,55,56], and future technologies are expected to reduce their influence further.

**Criterion for turn-on onset**

The absence of such criterion can lead to ambiguity in identifying the number and onset of turn-on events, potentially causing misinterpretation in situations involving crosstalk-induced gate spikes[25,38], multiple threshold crossings and multi-level gate-driving[57]. Within CUSST, the onset of a turn-on event is explicitly defined as the instant at which the initial rapid drop in $R_{S1}$ begins, triggered when $v_{gs,S1}$ exceeds the threshold voltage. Each threshold crossing that triggers such a decrease is counted as a distinct turn-on event. As the subsequent switching evolution originates from this initial rapid drop in $R_{S1}$, this instant also defines the starting point of the corresponding spatiotemporal evolution.

**Derivation of CUSST-based $E_{on}$ model in the representative iZVS scenario**

***Derivation from charge conservation***

With energy dissipation in $C_{S1}$ is negligible relative to dissipation along the $i_{RS1}$ paths[58,59], $E_{on,S1}$ is attributed entirely to $R_{S1}$:

$$E_{\text{on,S1}} = \int_{\text{switching-ON}} v_{\text{ds,S1}}(t) i_{\text{RS1}}(t) dt \,. \tag{1}$$

Applying KCL at $S_1$'s drain terminal, gives

$$i_{\text{RS1}}(t) = i_{\text{d,S1}}(t) + i_{C_{\text{gd,S1}}}(t) + i_{C_{\text{ds,S1}}}(t) + i_{\text{par},C_{\text{gd,S1}}}(t) + i_{\text{par},C_{\text{ds,S1}}}(t) \,. \tag{2}$$

Substituting (2) into (1), gives $E_{on,S1}$ as

$$E_{\text{on,S1}} = E_1 + E_2 + E_3 + E_4 + E_5 . \tag{3}$$

where

$$\left.\begin{aligned}
&E_1 = \int_{\text{switching-ON}} v_{\text{ds,S1}}(t) i_{\text{d,S1}}(t) dt, E_2 = \int_{\text{switching-ON}} v_{\text{ds,S1}}(t) i_{C_{\text{gd,S1}}}(t) dt, \\
&E_3 = \int_{\text{switching-ON}} v_{\text{ds,S1}}(t) i_{C_{\text{ds,S1}}}(t) dt, E_4 = \int_{\text{switching-ON}} v_{\text{ds,S1}}(t) i_{\text{par},C_{\text{gd,S1}}}(t) dt, \\
&E_5 = \int_{\text{switching-ON}} v_{\text{ds,S1}}(t) i_{\text{par},C_{\text{ds,S1}}}(t) dt,
\end{aligned}\right\} \tag{4}$$

Since $v_{gs,S1}$ during ($t_0$–$t_3$) is negligible relative to the residual drain-source voltage at $t_0$, denoted by $\Delta V$, $v_{gs,S1}$ can be approximated as the constant $V_{gp}$. Consequently, $\Delta V - V_{gp} \approx \Delta V$ and 0 V $- V_{gp} \approx$ 0 V. Consequently,

$$\left.\begin{aligned}
&E_2 + E_3 = E_{oss,S1}(\Delta V), \\
&E_4 + E_5 \approx \frac{1}{2} C_{par,S1} \Delta V^2,
\end{aligned}\right\} \tag{5}$$

where

$$\left.\begin{aligned}
E_4 &= \int_{0\mathrm{V}-V_{\mathrm{gp}}}^{\Delta V-V_{\mathrm{gp}}} v_{\mathrm{ds,S1}} C_{\mathrm{par,gd,S1}}\left(v_{\mathrm{gd,S1}}\right) dv_{\mathrm{gd,S1}} \approx \underbrace{\int_{0}^{\Delta V} v_{\mathrm{ds,S1}} C_{\mathrm{par,gd,S1}}\left(v_{\mathrm{gd,S1}}\right) dv_{\mathrm{gd,S1}}}_{\text{Energy dissipated by capacitance in parallel with } C_{\mathrm{gd,S1}}}, \\
E_5 &= \underbrace{\int_{0}^{\Delta V} v_{\mathrm{ds,S1}} C_{\mathrm{par,ds,S1}} dv_{\mathrm{ds,S1}}}_{\text{Energy dissipated by capacitance in parallel with } C_{\mathrm{ds,S1}}},
\end{aligned}\right\} \tag{6}$$

In addition, during ($t_0$–$t_3$), the charges transferred to $C_{\mathrm{gd,S2}}$, $C_{\mathrm{ds,S2}}$ and their corresponding parallel capacitances, $C_{\mathrm{par,gd,S2}}$ and $C_{\mathrm{par,ds,S2}}$, are denoted as $\Delta Q_{\mathrm{gd,S2}}$, $\Delta Q_{\mathrm{ds,S2}}$, $\Delta Q_{\mathrm{par,gd,S2}}$ and $\Delta Q_{\mathrm{par,ds,S2}}$, respectively. Denoting $\Delta Q_{\mathrm{C,S2}}$ as the total charge transferred to $C_{\mathrm{gd,S2}}$, $C_{\mathrm{ds,S2}}$ and their parallel capacitances, yields

$$\Delta Q_{C,S2} = \int_{switching-ON} i_{C,S2}\left(t\right) dt = \Delta Q_{S2} + C_{par,S2} \Delta V \tag{7}$$

where

$$\left.\begin{aligned}
\Delta Q_{\mathrm{S2}} &= \Delta Q_{\mathrm{gd,S2}} + \Delta Q_{\mathrm{ds,S2}} \approx Q_{\mathrm{oss,S2}}\left(V_{\mathrm{DC}}\right) - Q_{\mathrm{oss,S2}}\left(V_{\mathrm{DC}} - \Delta V\right), \\
\Delta Q_{\mathrm{par,S2}} &= \Delta Q_{\mathrm{par,gd,S2}} + \Delta Q_{\mathrm{par,ds,S2}} \approx C_{\mathrm{par,gd,S2}} \Delta V + C_{\mathrm{par,ds,S2}} \Delta V,
\end{aligned}\right\} \tag{8}$$

where

$$\left.\begin{aligned}
\Delta Q_{\mathrm{gd,S2}} &= \int_{\text{switching-ON}} i_{\mathrm{Cgd,S2}}\left(t\right) dt = \int_{V_{\mathrm{DC}}-\Delta V-(-V_{\mathrm{EE}})}^{V_{\mathrm{DC}}-(-V_{\mathrm{EE}})} C_{\mathrm{gd,S2}} dv_{\mathrm{gd,S2}} \\
&\approx \int_{V_{\mathrm{DC}}-\Delta V}^{V_{\mathrm{DC}}} C_{\mathrm{gd,S2}} dv_{\mathrm{gd,S2}} = \int_{0\mathrm{V}}^{V_{\mathrm{DC}}} C_{\mathrm{gd,S2}} dv_{\mathrm{gd,S2}} - \int_{0\mathrm{V}}^{V_{\mathrm{DC}}-\Delta V} C_{\mathrm{gd,S2}} dv_{\mathrm{gd,S2}} \\
&= Q_{\mathrm{gd,S2}}\left(V_{\mathrm{DC}}\right) - Q_{\mathrm{gd,S2}}\left(V_{\mathrm{DC}} - \Delta V\right), \\
\Delta Q_{\mathrm{ds,S2}} &= \int_{\text{switching-ON}} i_{\mathrm{Cds,S2}}\left(t\right) dt = \int_{V_{\mathrm{DC}}-\Delta V}^{V_{\mathrm{DC}}} C_{\mathrm{ds,S2}} dv_{\mathrm{ds,S2}} \\
&= \int_{V_{\mathrm{DC}}-\Delta V}^{V_{\mathrm{DC}}} C_{\mathrm{ds,S2}} dv_{\mathrm{ds,S2}} = \int_{0\mathrm{V}}^{V_{\mathrm{DC}}} C_{\mathrm{ds,S2}} dv_{\mathrm{ds,S2}} - \int_{0\mathrm{V}}^{V_{\mathrm{DC}}-\Delta V} C_{\mathrm{ds,S2}} dv_{\mathrm{ds,S2}} \\
&= Q_{\mathrm{ds,S2}}\left(V_{\mathrm{DC}}\right) - Q_{\mathrm{ds,S2}}\left(V_{\mathrm{DC}} - \Delta V\right), \\
\Delta Q_{\mathrm{par,gd,S2}} &= \int_{(V_{\mathrm{DC}}-\Delta V)-(-V_{\mathrm{EE}})}^{V_{\mathrm{DC}}-(-V_{\mathrm{EE}})} C_{\mathrm{par,gd,S2}}\left(v_{\mathrm{gd}}\right) dv_{\mathrm{gd}} \approx \int_{V_{\mathrm{DC}}-\Delta V}^{V_{\mathrm{DC}}} C_{\mathrm{par,gd,S2}}\left(v_{\mathrm{gd}}\right) dv_{\mathrm{gd}} \\
&= \left[\int_{0\mathrm{V}}^{V_{\mathrm{DC}}} C_{\mathrm{par,gd,S2}}\left(v_{\mathrm{gd}}\right) dv_{\mathrm{gd}} - \int_{0\mathrm{V}}^{V_{\mathrm{DC}}-\Delta V} C_{\mathrm{par,gd,S2}}\left(v_{\mathrm{gd}}\right) dv_{\mathrm{gd}}\right] = C_{\mathrm{par,gd,S2}} \Delta V, \\
\Delta Q_{\mathrm{par,ds,S2}} &= \int_{V_{\mathrm{DC}}-\Delta V}^{V_{\mathrm{DC}}} C_{\mathrm{par,ds,S2}}\left(v_{\mathrm{ds}}\right) dv_{\mathrm{ds}} = \int_{V_{\mathrm{DC}}-\Delta V}^{V_{\mathrm{DC}}} C_{\mathrm{par,ds,S2}}\left(v_{\mathrm{ds}}\right) dv_{\mathrm{ds}} \\
&= \left[\int_{0\mathrm{V}}^{V_{\mathrm{DC}}} C_{\mathrm{par,ds,S2}}\left(v_{\mathrm{ds}}\right) dv_{\mathrm{ds}} - \int_{0\mathrm{V}}^{V_{\mathrm{DC}}-\Delta V} C_{\mathrm{par,ds,S2}}\left(v_{\mathrm{ds}}\right) dv_{\mathrm{ds}}\right] = C_{\mathrm{par,ds,S2}} \Delta V,
\end{aligned}\right\} \tag{9}$$

Applying KVL across the loop incorporating the DC source, $S_1$ and $S_2$ yields

$$V_{\mathrm{DC}} = v_{\mathrm{ds,S1}}\left(t\right) + v_{\mathrm{ds,S2}}\left(t\right). \tag{10}$$

Applying KCL at the source terminal of $S_2$ yields

$$i_{\mathrm{DC}}\left(t\right) = i_{\mathrm{d,S2}}\left(t\right) - i_{\mathrm{L}}\left(t\right) = i_{\mathrm{RS2}}\left(t\right) + i_{\mathrm{C,S2}}\left(t\right) - i_{\mathrm{L}}\left(t\right). \tag{11}$$

Substituting (10) and (11) into the expression of $E_1$, yields

$$E_1 = E_{\mathrm{DC}} - E_{1,1} + E_{AC-link}, \tag{12}$$

where the energy transferred from the DC source, denoted by $E_{\mathrm{DC}}$, as well as two energy terms arising from breakdown of $E_1$, denoted by $E_{1,1}$ and $E_{\mathrm{AC\text{-}link}}$ are given by

$$\left.\begin{aligned}
E_{\mathrm{DC}} &= \int\limits_{\text{switching-ON}} V_{\mathrm{DC}} i_{\mathrm{DC}}(t)\,dt = V_{\mathrm{DC}} \int\limits_{\text{switching-ON}} \left[ i_{\mathrm{RS2}}(t) + i_{\mathrm{C,S2}}(t) - i_{\mathrm{L}}(t) \right] dt \\
&= V_{\mathrm{DC}} \left[ \int\limits_{\text{switching-ON}} i_{\mathrm{RS2}}(t)\,dt + \Delta Q_{S2} + C_{par,S2} \Delta V \right], \\
E_{1,1} &= \int\limits_{\text{switching-ON}} v_{\mathrm{ds,S2}}(t)\, i_{\mathrm{d,S2}}(t)\,dt = \Delta E_{\mathrm{gd,S2}} + \Delta E_{\mathrm{ds,S2}} + \Delta E_{\mathrm{par,gd,S2}} + \Delta E_{\mathrm{par,ds,S2}} + \Delta E_{\mathrm{RS2}}, \\
E_{AC-link} &= \int\limits_{\text{switching-ON}} v_{\mathrm{ds,S2}}(t)\, i_{\mathrm{L}}(t)\,dt,
\end{aligned}\right\} \tag{13}$$

where $\Delta E_{\mathrm{gd,S2}}$, $\Delta E_{\mathrm{ds,S2}}$, $\Delta E_{\mathrm{par,gd,S2}}$ and $\Delta E_{\mathrm{par,ds,S2}}$ denote transferred energies associated with $C_{\mathrm{gd,S2}}$, $C_{\mathrm{ds,S2}}$, and their parallel capacitances, respectively; $\Delta E_{\mathrm{RS2}}$ denotes the energy dissipation incurred in $R_{\mathrm{S2}}$ in case of a shoot-through event. These energy terms are given by

$$\left.\begin{aligned}
\Delta E_{\mathrm{gd,S2}} &= \int\limits_{(V_{\mathrm{DC}}-\Delta V)-(-V_{\mathrm{EE}})}^{V_{\mathrm{DC}}-(-V_{\mathrm{EE}})} v_{\mathrm{gd,S2}} C_{\mathrm{gd,S2}}\left(v_{\mathrm{gd,S2}}\right) dv_{\mathrm{gd,S2}} \approx \int\limits_{V_{\mathrm{DC}}-\Delta V}^{V_{\mathrm{DC}}} v_{\mathrm{gd,S2}} C_{\mathrm{gd,S2}}\left(v_{\mathrm{gd,S2}}\right) dv_{\mathrm{gd,S2}}, \\
\Delta E_{\mathrm{ds,S2}} &= \int\limits_{V_{\mathrm{DC}}-\Delta V}^{V_{\mathrm{DC}}} v_{\mathrm{ds,S2}} C_{\mathrm{ds,S2}}\left(v_{\mathrm{ds,S2}}\right) dv_{\mathrm{ds,S2}}, \\
\Delta E_{\mathrm{par,gd,S2}} &= \int\limits_{(V_{\mathrm{DC}}-\Delta V)-(-V_{\mathrm{EE}})}^{V_{\mathrm{DC}}-(-V_{\mathrm{EE}})} v_{\mathrm{gd,S2}} C_{\mathrm{par,gd,S2}}\left(v_{\mathrm{gd,S2}}\right) dv_{\mathrm{gd,S2}} \approx \int\limits_{V_{DC}-\Delta V}^{V_{DC}} v_{gd,S1} C_{par,gd,S2}\left(v_{gd,S2}\right) dv_{gd,S2}, \\
\Delta E_{\mathrm{par,ds,S2}} &= \int\limits_{V_{\mathrm{DC}}-\Delta V}^{V_{\mathrm{DC}}} v_{\mathrm{ds,S2}} C_{\mathrm{ds,S2}}\left(v_{\mathrm{ds,S2}}\right) dv_{\mathrm{ds,S2}}, \\
\Delta E_{\mathrm{RS2}} &= \int\limits_{switching-ON} v_{\mathrm{ds,S2}}(t)\, i_{\mathrm{RS2}}(t)\,dt,
\end{aligned}\right\} \tag{14}$$

Substituting (14) into (12) yields

$$\begin{aligned}
E_{1,1} \approx{}& \underbrace{E_{\mathrm{oss,S2}}(V_{\mathrm{DC}}) - E_{\mathrm{oss,S2}}(V_{\mathrm{DC}} - \Delta V)}_{\text{Energy stored by } C_{\mathrm{oss,S2}}} + \underbrace{\frac{1}{2} C_{\mathrm{par,S2}} \left[ V_{\mathrm{DC}}{}^2 - (V_{\mathrm{DC}} - \Delta V)^2 \right]}_{\text{Energy stored by the paralleled capacitance of } S2} \\
&+ \underbrace{\int\limits_{\text{switching-ON}} v_{\mathrm{ds,S2}}(t)\, i_{\mathrm{RS2}}(t)\,dt}_{\text{Energy dissipation in S2 due to shoot-through}}
\end{aligned} \tag{15}$$

Combining (1)(3)(4)(5)(8)(9)(12)(13)(14)(15) yields the $E_{\mathrm{on,S1}}$ model derived from the perspective of charge conservation:

$$E_{\mathrm{on},S1} = E_{DC} + E_{AC-link} - E_{1,1} + E_2 + E_3 + E_4 + E_5, \tag{16}$$

In detail, the $E_{\mathrm{on,S1}}$ model is expressed as follows, with interpretations provided for each energy term:

$$\begin{aligned} E_{\text{on,S1}} \approx & \underbrace{V_{DC}\left[\int\limits_{\text{switching-ON}} i_{\text{RS2}}(t)\,dt + \Delta Q_{\text{S2}} + C_{\text{par,S2}}\Delta V - \int\limits_{\text{switching-ON}} i_{\text{L}}(t)\,dt\right]}_{\text{Energy provided by DC source to the half-bridge}} \\ & + \underbrace{\int\limits_{\text{switching-ON}} v_{\text{ds,S2}}(t)\, i_L(t)\,dt}_{\text{Energy provided by the overall AC-link impedance to the half-bridge}} - \underbrace{\frac{1}{2}C_{\text{par,S2}}\left[{V_{\text{DC}}}^2 - (V_{\text{DC}} - \Delta V)^2\right]}_{\text{Energy stored by the paralleled capacitances of } S2} \\ & - \underbrace{\left[E_{\text{oss,S2}}(V_{\text{DC}}) - E_{\text{oss,S2}}(V_{\text{DC}} - \Delta V)\right]}_{\text{Energy stored by output capacitance of } S2} - \underbrace{\int\limits_{\text{switching-ON}} v_{\text{ds,S2}}(t)\, i_{\text{RS2}}(t)\,dt}_{\text{Energy dissipated in } R_{S2} \text{ due to shoot-through}} \\ & + \underbrace{E_{\text{oss,S1}}(\Delta V) + \frac{1}{2}C_{\text{par,S1}}\Delta V^2}_{\text{Energy dissipated in } R_{S1} \text{ due to discharge of } S1\text{'s output capacitance and paralleled capacitances}} . \end{aligned} \tag{17}$$

***Derivation from energy conservation***

During the $S_1$'s switching-on process, the half-bridge is taken as the study object. For simplicity reasons, the analysis is limited to ($t_0$–$t_3$) as vast majority of the energy dissipation during $S_1$'s turn-on is incurred during this interval. According to the law of energy conservation, the overall energy initially stored in the study object, minus the various dissipated energies incurred during the turn-on process and the energy delivered to the external circuit, yields the remaining stored energy at the end of the turn-on process. The general mathematical expression is given by

$$E_{\text{initial}} - E_{\text{dissipated}} - E_{\text{delivered}} = E_{\text{final}} \tag{18}$$

where $E_{\text{initial}}$ denotes the overall energy stored in the study object at the initial instant; $E_{\text{dissipated}}$ denotes the total energy losses during $S_1$'s turn-on process, including but not limited to turn-on loss and ESR losses; $E_{\text{delivered}}$ denotes the energy delivered to the external circuit; $E_{\text{final}}$ denotes the total energy stored in the study object at the end of the process. Among them, $E_{\text{initial}}$ is given by

$$E_{\text{initial}} = E_{\text{initial,S1}} + E_{\text{initial,S2}}, \tag{19}$$

where

$$\left.\begin{aligned} E_{\text{initial,S1}} &= E_{\text{initial,Cgd,S1}} + E_{\text{initial,Cds,S1}} + E_{\text{initial,par,gd,S1}} + E_{\text{initial,par,ds,S1}}, \\ E_{\text{initial,S2}} &= E_{\text{initial,Cgd,S2}} + E_{\text{initial,Cds,S2}} + E_{\text{initial,par,gd,S2}} + E_{\text{initial,par,ds,S2}}, \end{aligned}\right\} \tag{20}$$

where $E_{\text{initial,Sx}}$, $E_{\text{initial,Cgd,Sx}}$, $E_{\text{initial,Cds,Sx}}$, $E_{\text{initial,par,gd,Sx}}$, $E_{\text{initial,par,ds,Sx}}$ denote the total initial energy stored in the parasitic capacitances of $S_x$ (with x=1,2 representing the upper and lower devices, respectively), including the energies stored in $C_{\text{gd,Sx}}$, $C_{\text{ds,Sx}}$, as well as their corresponding parallel capacitances $C_{\mathrm{par,gd,Sx}}$ and $C_{\mathrm{par,gd,Sx}}$, respectively, given by:

$$\left.\begin{aligned}
&E_{\mathrm{initial,Cgd,S1}} = E_{\mathrm{gd,S1}}\left(\Delta V - v_{\mathrm{gs,S1}}\left(t_0\right)\right), E_{\mathrm{initial,Cds,S1}} = E_{\mathrm{ds,S1}}\left(\Delta V\right),\\
&E_{\mathrm{initial,par,gd,S1}} = \frac{1}{2}C_{\mathrm{par,gd,S1}}\left(\Delta V - v_{\mathrm{gs,S1}}\left(t_0\right)\right)^2, E_{\mathrm{initial,par,ds,S1}} = \frac{1}{2}C_{\mathrm{par,ds,S1}}\left(\Delta V\right)^2,\\
&E_{\mathrm{initial,Cgd,S2}} = E_{\mathrm{gd,S2}}\left(V_{\mathrm{DC}} - \Delta V - v_{\mathrm{gs,S2}}\left(t_0\right)\right), E_{\mathrm{initial,Cds,S2}} = E_{\mathrm{ds,S2}}\left(V_{\mathrm{DC}} - \Delta V\right),\\
&E_{\mathrm{initial,par,gd,S2}} = \frac{1}{2}C_{\mathrm{par,gd,S2}}\left(V_{\mathrm{DC}} - \Delta V - v_{\mathrm{gs,S2}}\left(t_0\right)\right)^2,\\
&E_{\mathrm{initial,par,ds,S2}} = \frac{1}{2}C_{\mathrm{par,ds,S2}}\left(V_{\mathrm{DC}} - \Delta V\right)^2,
\end{aligned}\right\} \tag{21}$$

where $v_{\mathrm{gs,S1}}(t_0)=V_{\mathrm{th,S1}}$.

Similarly, $E_{\mathrm{final}}$ is given by

$$E_{\mathrm{final}}=E_{\mathrm{final,S1}}+E_{\mathrm{final,S2}}, \tag{22}$$

where

$$\left.\begin{aligned}
E_{\mathrm{final,S1}} &= E_{\mathrm{final,Cgd,S1}} + E_{\mathrm{final,Cds,S1}} + E_{\mathrm{final,par,gd,S1}} + E_{\mathrm{final,par,ds,S1}},\\
E_{\mathrm{final,S2}} &= E_{\mathrm{final,Cgd,S2}} + E_{\mathrm{final,Cds,S2}} + E_{\mathrm{final,par,gd,S2}} + E_{\mathrm{final,par,ds,S2}},
\end{aligned}\right\} \tag{23}$$

where

$$\left.\begin{aligned}
&E_{\mathrm{final,Cgd,S1}} = E_{\mathrm{gd,S1}}\left(0\mathrm{V} - v_{\mathrm{gs,S1}}\left(t_3\right)\right), E_{\mathrm{final,Cds,S1}} = E_{\mathrm{ds,S1}}\left(0\mathrm{V}\right),\\
&E_{\mathrm{final,par,gd,S1}} = \frac{1}{2}C_{\mathrm{par,gd,S1}}\left(0\mathrm{V} - v_{\mathrm{gs,S1}}\left(t_3\right)\right)^2, E_{\mathrm{final,par,ds,S1}} = \frac{1}{2}C_{\mathrm{par,ds,S1}}\left(0\mathrm{V}\right)^2,\\
&E_{\mathrm{final,Cgd,S2}} = E_{\mathrm{gd,S2}}\left(V_{\mathrm{DC}} - 0\mathrm{V} - v_{\mathrm{gs,S2}}\left(t_3\right)\right), E_{\mathrm{final,Cds,S2}} = E_{\mathrm{ds,S2}}\left(V_{\mathrm{DC}} - 0\mathrm{V}\right),\\
&E_{\mathrm{final,par,gd,S2}} = \frac{1}{2}C_{\mathrm{par,gd,S2}}\left(V_{\mathrm{DC}} - 0\mathrm{V} - v_{\mathrm{gs,S2}}\left(t_3\right)\right)^2, E_{\mathrm{final,par,ds,S2}} = \frac{1}{2}C_{\mathrm{par,ds,S2}}\left(V_{\mathrm{DC}} - 0\mathrm{V}\right)^2,
\end{aligned}\right\} \tag{24}$$

Since both $v_{\mathrm{gs,S1}}$ and $v_{\mathrm{gs,S2}}$ at any instant during the switching-on process are negligible relative to $\Delta V$ or $V_{DC}-\Delta V$, both $v_{\mathrm{gs,S1}}$ and $v_{\mathrm{gs,S2}}$ can be approximated to 0 V, which yields

$$\left.\begin{aligned}
&E_{\mathrm{initial,Cgd,S1}} \approx E_{\mathrm{gd,S1}}\left(\Delta V\right), E_{\mathrm{initial,Cds,S1}} = E_{\mathrm{ds,S1}}\left(\Delta V\right),\\
&E_{\mathrm{initial,par,gd,S1}} \approx \frac{1}{2}C_{\mathrm{par,gd,S1}}\left(\Delta V\right)^2, E_{\mathrm{initial,par,ds,S1}} = \frac{1}{2}C_{\mathrm{par,ds,S1}}\left(\Delta V\right)^2,\\
&E_{\mathrm{initial,Cgd,S2}} \approx E_{\mathrm{gd,S2}}\left(V_{\mathrm{DC}} - \Delta V\right), E_{\mathrm{initial,Cds,S2}} = E_{\mathrm{ds,S2}}\left(V_{\mathrm{DC}} - \Delta V\right),\\
&E_{\mathrm{initial,par,gd,S2}} \approx \frac{1}{2}C_{\mathrm{par,gd,S2}}\left(V_{\mathrm{DC}} - \Delta V\right)^2, E_{\mathrm{initial,par,ds,S2}} = \frac{1}{2}C_{\mathrm{par,ds,S2}}\left(V_{\mathrm{DC}} - \Delta V\right)^2,\\
&E_{\mathrm{final,Cgd,S1}} \approx E_{\mathrm{gd,S1}}\left(0\mathrm{V}\right) = 0\mathrm{J}, E_{\mathrm{final,Cds,S1}} = E_{\mathrm{ds,S1}}\left(0\mathrm{V}\right) = 0\mathrm{J},\\
&E_{\mathrm{final,par,gd,S1}} \approx \frac{1}{2}C_{\mathrm{par,gd,S1}}\left(0\mathrm{V}\right)^2 = 0\mathrm{J}, E_{\mathrm{final,par,ds,S1}} = \frac{1}{2}C_{\mathrm{par,ds,S1}}\left(0\mathrm{V}\right)^2 = 0\mathrm{J},\\
&E_{\mathrm{final,Cgd,S2}} \approx E_{\mathrm{gd,S2}}\left(V_{\mathrm{DC}} - 0\mathrm{V}\right), E_{\mathrm{final,Cds,S2}} = E_{\mathrm{ds,S2}}\left(V_{\mathrm{DC}} - 0\mathrm{V}\right),\\
&E_{\mathrm{final,par,gd,S2}} \approx \frac{1}{2}C_{\mathrm{par,gd,S2}}\left(V_{\mathrm{DC}} - 0\mathrm{V}\right)^2, E_{\mathrm{final,par,ds,S2}} = \frac{1}{2}C_{\mathrm{par,ds,S2}}\left(V_{\mathrm{DC}} - 0\mathrm{V}\right)^2,
\end{aligned}\right\} \tag{25}$$

Consequently, $E_{\mathrm{initial}}$ and $E_{\mathrm{final}}$ can be approximated as

$$\left.\begin{aligned} E_{\text{initial}} &\approx \left[ E_{\text{oss,S1}}\left(\Delta V\right)+\frac{1}{2}C_{\text{par,S1}}\left(\Delta V\right)^{2} \right]+\left[ E_{\text{oss,S2}}\left(V_{\text{DC}}-\Delta V\right)+\frac{1}{2}C_{\text{par,S2}}\left(V_{DC}-\Delta V\right)^{2} \right], \\ E_{\text{final}} &\approx E_{\text{oss,S2}}\left(V_{\text{DC}}\right)+\frac{1}{2}C_{\text{par,S2}}V_{\text{DC}}{}^{2}. \end{aligned}\right\} \tag{26}$$

During $S_1$'s turn-on process, the dominant energy dissipation in the study object is $E_{\text{on,S1}}$ and the dissipation incurred within $R_{\text{S2}}$ in the case of a shoot-through event. Hence, $E_{\text{dissipated}}$ can be approximated as comprising only these two components, namely

$$E_{\text{dissipated}} \approx E_{\text{on,S1}} + E_{\text{dissipated,S2}}, \tag{27}$$

where

$$E_{\text{dissipated,S2}} = \underbrace{\int_{\text{switching-ON}} v_{\text{ds,S2}}\left(t\right)i_{\text{RS2}}\left(t\right)dt}_{\text{Energy dissipated in Rs2 in the event of shoot through}}. \tag{28}$$

The energy delivered to the external circuit by the study obeject, namely $E_{\text{delivered}}$ could be derived as

$$E_{\text{delivered}} = -\left(W_{\text{DC}} + W_{\text{L}}\right), \tag{29}$$

where $W_{\text{DC}}$ and $W_{\text{L}}$ denote the work done by the DC source and load inductor to the study object, respectively. In order to obtain $W_{\text{DC}}$, it is important to determine the total charge transferred by the DC source, denoted as $\Delta Q_{\text{DC}}$ over the turn-on process, while the charges transferred to $C_{\text{gd,S2}}$, $C_{\text{ds,S2}}$ and their corresponding parallel capacitances are denoted as $\Delta Q_{\text{gd,S2}}$, $\Delta Q_{\text{ds,S2}}$, $\Delta Q_{\text{par,gd,S2}}$ and $\Delta Q_{\text{par,ds,S2}}$, respectively, together with their sum denoted as $\Delta Q_{\text{C,S2}}$; $\Delta Q_{\text{RS2}}$ and $\Delta Q_{\text{L}}$ denote the charge through $R_{\text{S2}}$ and load inductor, respectively.

Applying KCL at $S_2$'s source terminal yields

$$i_{\text{DC}}\left(t\right) = i_{\text{d,S2}}\left(t\right) - i_{\text{L}}\left(t\right) = i_{\text{C,S2}}\left(t\right) + i_{\text{RS2}}\left(t\right) - i_{\text{L}}\left(t\right). \tag{30}$$

Integrating both sides of (30) yields

$$\Delta Q_{\text{DC}} = \Delta Q_{\text{C,S2}} + \Delta Q_{\text{RS2}} - \Delta Q_{\text{L}}, \tag{31}$$

where

$$\left.\begin{aligned} \Delta Q_{\text{DC}} &= \int_{\text{switching-ON}} i_{\text{DC}}\left(t\right)dt, \Delta Q_{\text{C,S2}} = \int_{\text{switching-ON}} i_{\text{C,S2}}\left(t\right)dt, \\ \Delta Q_{\text{RS2}} &= \int_{\text{switching-ON}} i_{\text{RS2}}\left(t\right)dt, \Delta Q_{\text{L}} = \int_{\text{switching-ON}} i_{\text{L}}\left(t\right)dt, \end{aligned}\right\} \tag{32}$$

where

$$\Delta Q_{\text{C,S2}} = \Delta Q_{\text{gd,S2}} + \Delta Q_{\text{ds,S2}} + \Delta Q_{\text{par,gd,S2}} + \Delta Q_{\text{par,ds,S2}}, \tag{33}$$

where

$$
\left.
\begin{aligned}
&\Delta Q_{\mathrm{gd,S2}} = \int\limits_{\text{switching-ON}} i_{\mathrm{Cgd,S2}}\left(t\right)dt = \int\limits_{\text{switching-ON}} C_{\mathrm{gd,S2}}\frac{dv_{\mathrm{gd,S2}}}{dt}dt \\
&= \int_{V_{\mathrm{DC}}-\Delta V - v_{\mathrm{gs,S2}}(t_0)}^{V_{\mathrm{DC}}-v_{\mathrm{gs,S2}}(t_3)} C_{\mathrm{gd,S2}}dv_{\mathrm{gd,S2}} \approx \int_{V_{\mathrm{DC}}-\Delta V}^{V_{\mathrm{DC}}} C_{\mathrm{gd,S2}}dv_{\mathrm{gd,S2}}, \\
&\Delta Q_{\mathrm{ds,S2}} = \int\limits_{\text{switching-ON}} i_{\mathrm{Cds,S2}}\left(t\right)dt = \int\limits_{\text{switching-ON}} C_{\mathrm{ds,S2}}\frac{dv_{\mathrm{ds,S2}}}{dt}dt \\
&= \int_{V_{\mathrm{DC}}-\Delta V}^{V_{\mathrm{DC}}} C_{\mathrm{ds,S2}}dv_{\mathrm{ds,S2}} \approx \int_{V_{\mathrm{DC}}-\Delta V}^{V_{\mathrm{DC}}} C_{\mathrm{ds,S2}}dv_{\mathrm{ds,S2}}, \\
&\Delta Q_{\mathrm{par,gd,S2}} = \int\limits_{\text{switching-ON}} i_{\mathrm{Cpar,gd,S2}}\left(t\right)dt = \int\limits_{\text{switching-ON}} C_{\mathrm{par,gd,S2}}\frac{dv_{\mathrm{gd,S2}}}{dt}dt \\
&= \int_{V_{\mathrm{DC}}-\Delta V - v_{\mathrm{gs,S2}}(t_0)}^{V_{\mathrm{DC}}-v_{\mathrm{gs,S2}}(t_3)} C_{\mathrm{par,gd,S2}}dv_{\mathrm{gd,S2}} \approx \int_{V_{\mathrm{DC}}-\Delta V}^{V_{\mathrm{DC}}} C_{\mathrm{par,gd,S2}}dv_{\mathrm{gd,S2}}, \\
&\Delta Q_{\mathrm{par,ds,S2}} = \int\limits_{\text{switching-ON}} i_{\mathrm{Cpar,ds,S2}}\left(t\right)dt = \int\limits_{\text{switching-ON}} C_{\mathrm{par,ds,S2}}\frac{dv_{\mathrm{ds,S2}}}{dt}dt \\
&= \int_{V_{\mathrm{DC}}-\Delta V}^{V_{\mathrm{DC}}} C_{\mathrm{par,ds,S2}}dv_{\mathrm{ds,S2}} \approx \int_{V_{\mathrm{DC}}-\Delta V}^{V_{\mathrm{DC}}} C_{\mathrm{par,ds,S2}}dv_{\mathrm{ds,S2}},
\end{aligned}
\right\} \tag{34}
$$

Consequently, $\Delta Q_{\mathrm{C,S2}}$ can be approximated as

$$
\Delta Q_{\mathrm{C,S2}} \approx Q_{\mathrm{oss,S2}}\left(V_{\mathrm{DC}}\right) - Q_{\mathrm{oss,S2}}\left(V_{\mathrm{DC}}-\Delta V\right) + \left(C_{\mathrm{par,gd,S2}} + C_{\mathrm{par,ds,S2}}\right)\Delta V = \Delta Q_{\mathrm{S2}} + C_{\mathrm{par,S2}}\Delta V, \tag{35}
$$

Hence, $W_{\mathrm{DC}}$ can be determined as

$$
\begin{aligned}
W_{\mathrm{DC}} &= \int\limits_{\text{switching-ON}} V_{\mathrm{DC}} i_{\mathrm{DC}}\left(t\right)dt = V_{\mathrm{DC}}\Delta Q_{\mathrm{DC}} \\
&= V_{\mathrm{DC}}\left[\Delta Q_{\mathrm{S2}} + C_{\mathrm{par,S2}}\Delta V + \int\limits_{\text{switching-ON}} i_{\mathrm{RS2}}\left(t\right)dt - \int\limits_{\text{switching-ON}} i_{\mathrm{L}}\left(t\right)dt\right].
\end{aligned} \tag{36}
$$

The work done by the load inductor, denoted by $W_{\mathrm{L}}$ can be determined as

$$
W_{\mathrm{L}} = \int\limits_{\text{switching-ON}} v_{\mathrm{L}}\left(t\right)i_{\mathrm{L}}\left(t\right)dt = \int\limits_{\text{switching-ON}} v_{\mathrm{ds,S2}}\left(t\right)i_{\mathrm{L}}\left(t\right)dt\,. \tag{37}
$$

Substituting (36) and (37) into (29) yields

$$
\begin{aligned}
&E_{\mathrm{delivered}} = -\left(W_{\mathrm{DC}} + W_{\mathrm{L}}\right) \\
&= -V_{\mathrm{DC}}\left[\Delta Q_{\mathrm{S2}} + C_{\mathrm{par,S2}}\Delta V + \int\limits_{\text{switching-ON}} i_{\mathrm{RS2}}\left(t\right)dt - \int\limits_{\text{switching-ON}} i_{\mathrm{L}}\left(t\right)dt\right] - \int\limits_{\text{switching-ON}} v_{\mathrm{ds,S2}}\left(t\right)i_{\mathrm{L}}\left(t\right)dt.
\end{aligned} \tag{38}
$$

Combining (26)(27)(28)(38) yields the $E_{\mathrm{on,S1}}$ model from perspective of energy conservation:

$$
\begin{aligned}
E_{\mathrm{on,S1}} \approx\ & \underbrace{V_{\mathrm{DC}}\left[\int\limits_{\text{switching-ON}} i_{\mathrm{RS2}}(t)dt + \Delta Q_{\mathrm{S2}} + C_{\mathrm{par,S2}}\Delta V - \int\limits_{\text{switching-ON}} i_{\mathrm{L}}(t)dt\right]}_{\text{Energy provided by DC source to the half-bridge}} \\
& + \underbrace{\int\limits_{\text{switching-ON}} v_{\mathrm{ds,S2}}(t)\, i_{\mathrm{L}}(t)dt}_{\text{Energy provided by the load inductor to the half-bridge}} - \underbrace{\left[E_{\mathrm{oss,S2}}\left(V_{\mathrm{DC}}\right) - E_{\mathrm{oss,S2}}\left(V_{\mathrm{DC}} - \Delta V\right)\right]}_{\text{Energy stored by output capacitance of } S_2} \\
& - \underbrace{\int\limits_{\text{switching-ON}} v_{\mathrm{ds,S2}}(t)\, i_{\mathrm{RS2}}(t)dt}_{\text{Energy dissipated in } S_2 \text{ due to shoot-through}} - \underbrace{\frac{1}{2}C_{\mathrm{par,S2}}\left[{V_{\mathrm{DC}}}^2 - \left(V_{\mathrm{DC}} - \Delta V\right)^2\right]}_{\text{Energy stored by the paralleled capacitance of } S_2} \\
& + \underbrace{E_{\mathrm{oss,S1}}\left(\Delta V\right) + \frac{1}{2}C_{\mathrm{par,S1}}\Delta V^2}_{\text{Energy dissipated in } R_{S1} \text{ due to discharge of } S_1\text{'s output capacitance and paralleled capacitance}}
\end{aligned}
\tag{39}
$$

Notably, (39) is identical to (17), which demonstrates that the energy- and charge-conservation derivations yield the same CUSST-based $E_{\mathrm{on}}$ model. This confirms the equivalent applicability of both conservation laws to semiconductor switching and extends their applicability to semiconductor devices. The equivalence further reveals two complementary physical views in the switching dynamics: energy conservation highlights the net energy redistribution and dissipation, while charge conservation elucidates the spatiotemporal charge evolution.

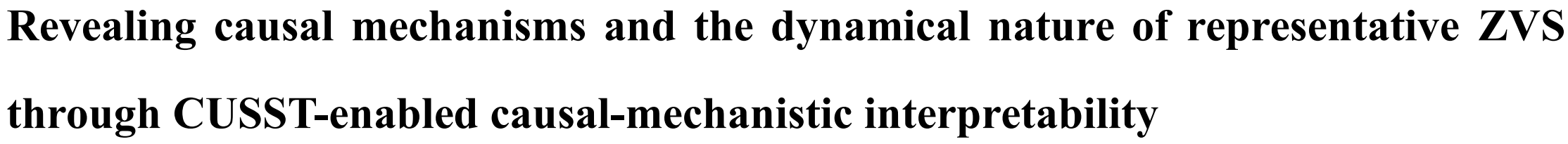

### Revealing causal mechanisms and the dynamical nature of representative ZVS through CUSST-enabled causal-mechanistic interpretability

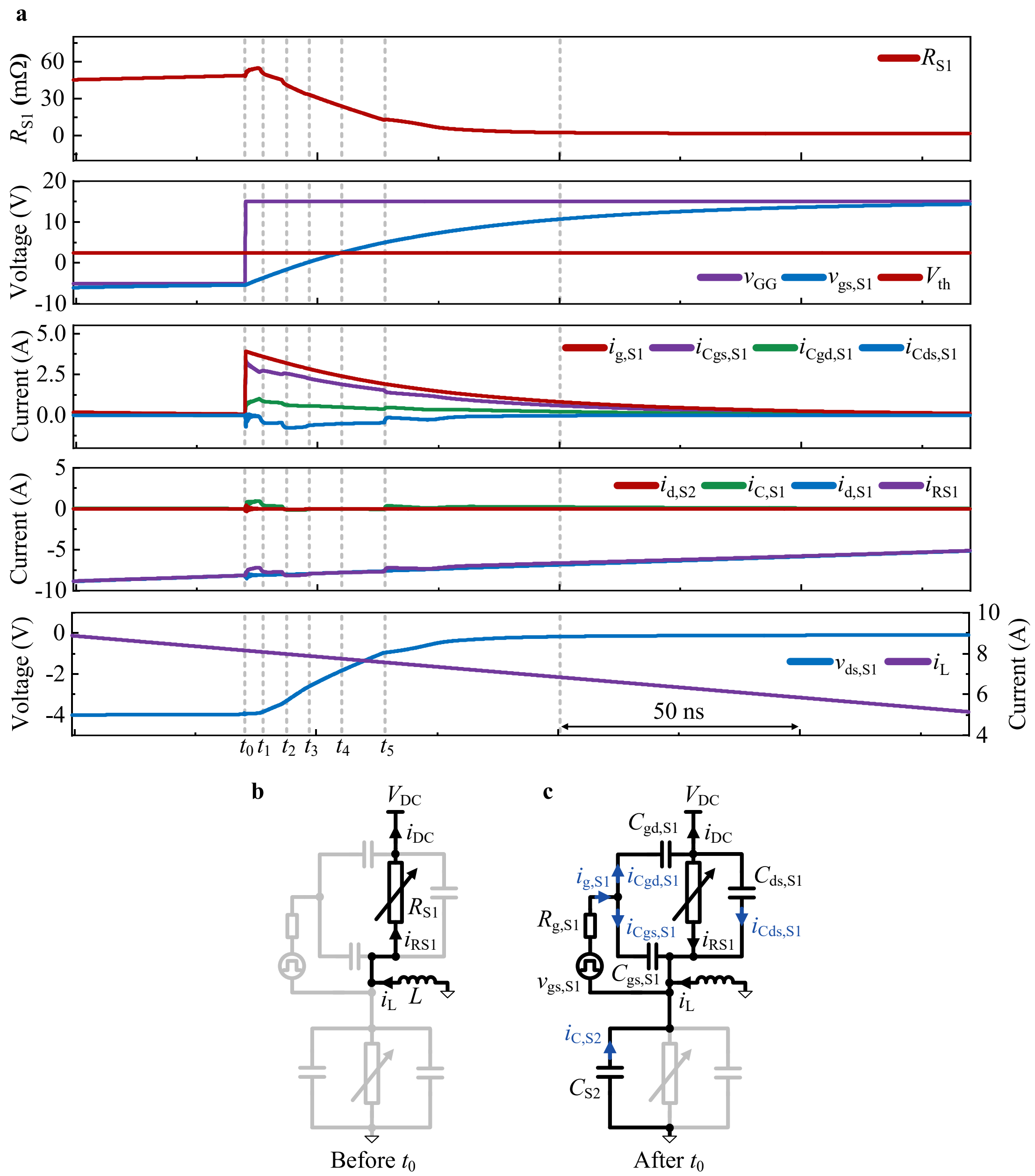

**Fig. 5| Revealing causal mechanisms and the dynamical nature of representative ZVS through CUSST-enabled causal-mechanistic interpretability. a**, Switching waveforms during a representative ZVS, obtained using LTspice. **b**, Operation mode before $t_0$. **c**, Operation mode after $t_0$. The equivalent-circuit formalism is obtained by $\mathbf{P_1}$**;** the highlighted branches in **b-c** indicate the current paths; the current components that differ from those in the immediately preceding sub-figure are highlighted in blue.

Before $t_0$, $i_L$ is reverse-conducted by $R_{S1}$ and gradually decreases, while its low decay rate has negligible impact on $v_{ds,S1}$. During ($t_0$–$t_1$), since $v_{gs,S1}$<–4 V, its rise has negligible influence on $R_{S1}$. As $i_{ds,S1}$ decreases, $v_{ds,S1}$ decreases correspondingly (by $\mathbf{P_2}$ and Ohm's law, with third-quadrant characteristics considered). Meanwhile, $i_{g,S1}$ simultaneously charges $C_{gs,S1}$ and discharges $C_{gd,S1}$. Applying KCL at the half-bridge midpoint yields $|i_{RS1}|+|i_{Cgd,S1}|=|i_L|+|i_{Cds,S1}|$. Since $|i_{Cgd,S1}|>|i_{Cds,S1}|$, it follows that $|i_{RS1}|<|i_L|$, which explains the brief drop of $|i_{RS1}|$ relative to $|i_L|$, and consequently causes a brief rise in $R_{S1}$ (by **P2**, with third-quadrant characteristics considered). As $C_{ds,S1}$ is high and $i_{Cds,S1}$ remains low, the resulting $S_1$'s *dv/dt* is limited, which keeps $v_{ds,S1}$ nearly constant as a result.

After $t_1$, $v_{gs,S1}>-4$ V. The rise in $v_{gs,S1}$ significantly reduces $R_{S1}$, which in turn lowers $v_{ds,S1}$ (by **P2** and Ohm's law; with third-quadrant characteristics considered). As a result, both $C_{ds,S1}$ and $C_{S2}$ discharge via $R_{S1}$. During ($t_2$–$t_3$), $R_{S1}$ drops more rapidly, which produces a steeper decline in $v_{ds,S1}$ and hence a higher $i_{Cds,S1}$ to discharge $C_{ds,S1}$. During ($t_1$–$t_2$) and ($t_3$–$t_4$), $R_{S1}$ decreases more slowly, which leads to a gentler drop in $v_{ds,S1}$ and a relatively lower $i_{Cds,S1}$.

After $t_4$, the voltage across the junction across p-body/drift region falls below the built-in voltage, which suppresses carrier injection from the p-body. A recovery current $i_{rr,S1}$, then arises within $S_1$ to reduce the plasma stored in the drift region. During ($t_4$–$t_5$), as $v_{gs,S1}<V_{th,S1}$, which keeps $S_1$'s channel off; thus, both $i_{RS1}$ and $i_{rr,S1}$ are entirely conducted by the diode. Because of the significant increase of the overall conduction current given by $i_{RS1}+i_{rr,S1}$, $v_{ds,S1}$ stops decreasing and nearly stabilizes, and even causes a brief rise near $t_5$.

During ($t_5$–$t_6$), $v_{gs,S1}>V_{th,S1}$, which turns on $S_1$'s channel; thus, the overall conduction current $i_{RS1}+i_{rr,S1}$ commutates into $S_1$'s channel as the channel resistance decreases gradually, and $v_{ds,S1}$ resumes its decline after the brief rise near $t_5$. After $t_6$, the commutation continues towards $S_1$'s channel. From $t_7$, $i_{rr,S1}$ begins to drop, which slows the reduction of $R_{S1}$ due to the rise of $v_{gs,S1}$. $i_{RS1}+i_{rr,S1}$ continues to commutate to the channel until $t_8$, when the entire $i_{RS1}+i_{rr,S1}$ is conducted by the channel.

After $t_8$, the rise $v_{gs,S1}$ has less influence on $R_{S1}$. As recovery current decreases, the half-bridge midpoint voltage drops slightly, with minor discharges of $C_{oss,S1}$ and $C_{S2}$ via $R_{S1}$ simultaneously.

## Revealing causal mechanisms and the dynamical nature of another iZVS through CUSST-enabled causal-mechanistic interpretability

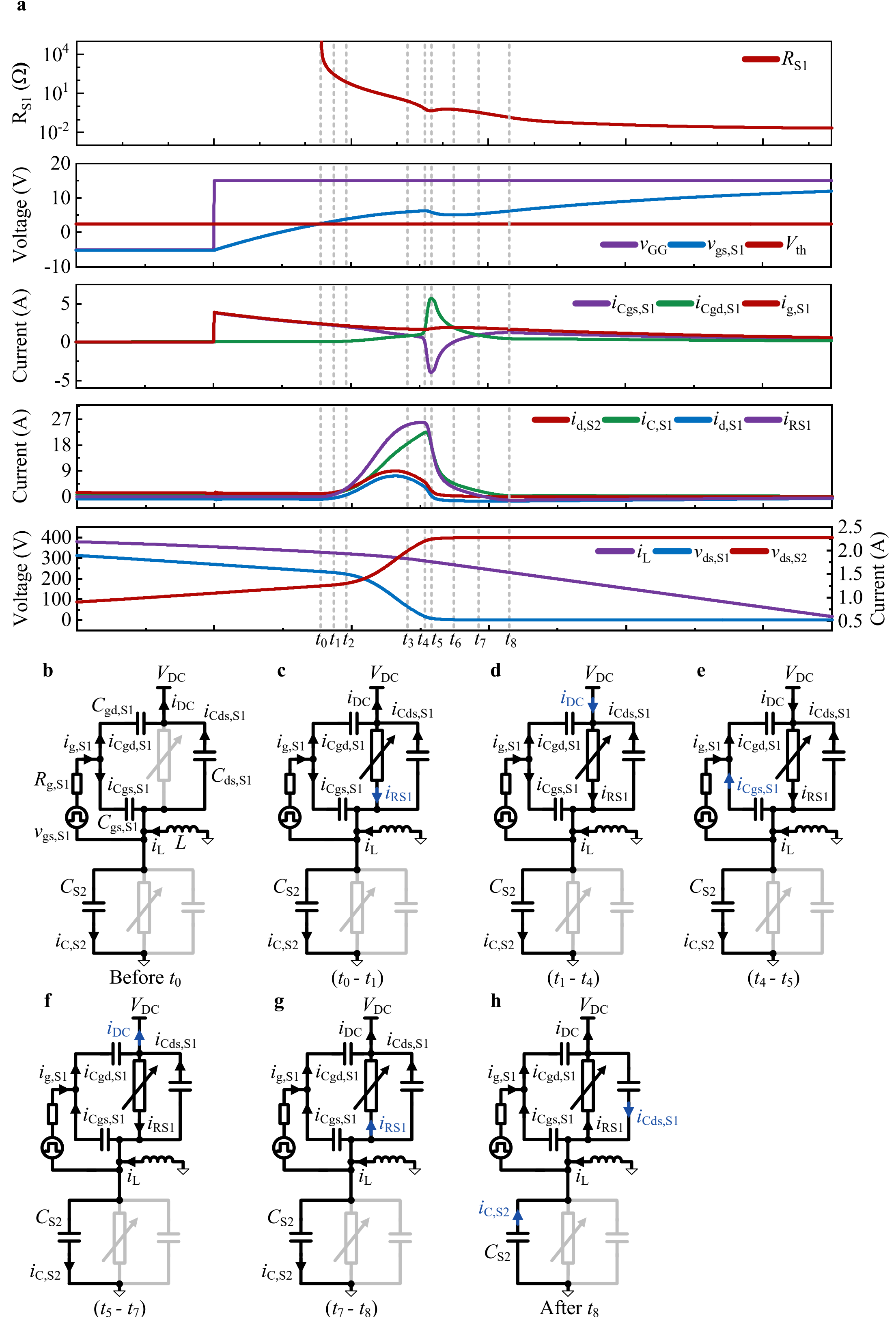

**Fig. 6 | Causal-mechanism interpretation for switching waveforms within another iZVS scenario.**

**a**, Switching waveforms within another iZVS scenario ($i_L$ flows into the half-bridge midpoint throughout the entire iZVS process). **b**, Operation mode before $t_0$. **c**, Operation mode during subinterval ($t_0$-$t_1$). **d**, Operation mode during subinterval ($t_1$-$t_4$). **e**, Operation mode during subinterval ($t_4$-$t_5$). **f**, Operation mode during subinterval ($t_5$-$t_7$). **g**, Operation mode during subinterval ($t_7$-$t_8$). **h**, Operation mode after $t_8$. The equivalent-circuit formalism is obtained by **P**$_1$**;** the highlighted branches in **b-h** indicate the current paths; the current components that differ from those in the immediately preceding sub-figure are highlighted in blue.

Before $t_0$ (i.e., the onset of iZVS), $i_L$ discharges $C_{S1}$ and charges $C_{S2}$ simultaneously. During ($t_0$-$t_1$), the low $i_{d,S2}$ results in a low $S_2$'s *dv/dt*, despite a relatively low $C_{oss,S2}$ (by **P4**; considering $C_{oss,S2}$'s nonlinearity; due to the relatively high $v_{ds,S2}$ compared to that in HS). Consequently, only a small portion of $i_{g,S1}$ discharging $C_{gd,S1}$ is required to follow $S_2$'s *dv/dt*, allowing most of $i_{g,S1}$ to charge $C_{gs,S1}$. This leads to a significant increase in $v_{gs,S1}$, sharply reducing $R_{S1}$ (by **P**$_3$), consequently increasing $i_{RS1}$ significantly. As a result, a lossy CC occurs, where the $C_{S1}$-conducted share of $i_L$ commutates to $C_{S2}$ (by **P4**; accounting for influence of $i_L$). Unlike the near-zero $S_2$'s *dv/dt* in the CC of HS, this CC features a non-zero *dv/dt* that increases with time, due to the increasing $i_{CS2}$ and decreasing $C_{oss,S2}$ (by **P4**; considering $C_{oss,S2}$'s nonlinearity). At $t_1$, $i_{RS1}$ exceeds $i_{C,S1}$, indicating the completion of the CC and triggering a reversal of $i_{DC}$'s direction.

During ($t_1$-$t_2$), $i_{RS1}+i_L=i_{d,S2}+i_{C,S1}$ (by **P4**; considering influences of other elements including $i_L$). Due to the initially high $R_{S1}$ (by **P**$_3$), $i_{RS1}$ and thus $i_{d,S2}$ are low. As $v_{gs,S1}$ increases, $R_{S1}$ decreases (by **P**$_3$), increasing $i_{RS1}$ and consequently increasing $i_{d,S2}$. Combined with the much lower $C_{oss,S2}$, increasing $i_{d,S2}$ leads to a higher $S_2$'s *dv/dt* compared to that in HS. The higher $S_2$'s *dv/dt*, combined with the higher $C_{oss,S1}$ (by **P4**; considering $C_{oss,S1}$'s nonlinearity), results in a larger $i_{C,S1}$. This, in turn, limits $i_{d,S2}$'s increase, thereby slowing *dv/dt* rise. Consequently, only a small portion of $i_{g,S1}$ is diverted to discharge $C_{gd,S1}$ to follow $S_2$'s *dv/dt*, while the majority of $i_{g,S1}$ charges $C_{gs,S1}$.

During ($t_2$-$t_3$), the continued increase in $i_{RS1}$ leads to an increase in $i_{d,S2}$, consequently an increase in $S_2$'s *dv/dt*. As negative feedback, more $i_{g,S1}$ is diverted to $C_{gd,S1}$, leading to two consequences: (1) higher $i_{Cgd,S1}$, promoting $C_{gd,S1}$'s *dv/dt*; (2) lower $i_{Cgs,S1}$, which slows the increase in $v_{gs,S1}$, thereby slowing $R_{S1}$ reduction (by **P**$_3$) and consequently slowing $i_{RS1}$ increase. The slower increase in $i_{RS1}$, combined with a significant increase in $i_{C,S1}$, reduces the *di/dt* of $i_{d,S2}$ – initially positive in ($t_2$-$t_3$), eventually becoming negative before $t_3$ - limiting increase in $S_2$'s $dv/dt$ despite the decreasing $C_{oss,S2}$ (by **P4**; considering $C_{oss,S2}$'s nonlinearity). As a result of these combined effects, $C_{gd,S1}$'s $dv/dt$ follows $S_2$'s $dv/dt$.

During ($t_3$-$t_4$), the quicker relative reduction in $v_{ds,S1}$ than the relative reduction in $R_{S1}$ (by **P3**), causes a slight decrease in $i_{RS1}$, contributing to a decreasing $i_{d,S2}$. A significant increase in $C_{gd,S1}$ and $C_{ds,S1}$ causes an increase in $i_{Cgd,S1}$ and $i_{Cds,S1,}$ respectively, and thus the increase in $i_{C,S1}$, also contributing to the decrease in $i_{d,S2}$. Despite a slower relative drop in $C_{oss,S2}$ (by **P4**; considering $C_{oss,S2}$'s nonlinearity), the significant decrease in $i_{d,S2}$ leads to a slight decrease in $S_2$'s $dv/dt$ and consequently a slight decrease in $S_1$'s $dv/dt$. Hence, more $i_{g,S1}$ is diverted to discharge $C_{gd,S1}$, further slowing the increase of $v_{gs,S1}$ and thus slowing the reduction in $R_{S1}$ (by **P3**).

During ($t_4$-$t_6$), $C_{gd,S1}$ is in its high-capacitance region (by **P4**; considering $C_{gd,S1}$'s nonlinearity), increasing rapidly as $v_{ds,S1}$ decreases; as $v_{ds,S2}$ approaches $V_{DC}$, $C_{oss,S2}$ remains low (by **P4**; considering $C_{oss,S2}$'s nonlinearity); as $i_{d,S2}$ remains significant, $S_2$'s $dv/dt$ remains significant. Consequently, $C_{gd,S1}$'s $dv/dt$ has to remain significant. The insufficient $i_{g,S1}$ to maintain $C_{gd,S1}$'s $dv/dt$ to follow $S_2$'s $dv/dt$, triggers further negative feedback – a drop in $v_{gs,S1}$. This boosts $i_{g,S1}$, enhancing $i_{Cgd,S1}$, but also increases $R_{S1}$ (by **P3**), which combined with decreasing $v_{ds,S1}$, leads to a rapid reduction in $i_{RS1}$, and consequently a decreasing $i_{d,S2}$, thereby a lower $dv/dt$. Meanwhile, $C_{gs,S1}$ supplies a pulse current to help to discharge $C_{gd,S1}$ via $R_{S1}$. Together, these effects enable $C_{gd,S1}$'s $dv/dt$ to follow $S_2$'s $dv/dt$. Notably, at $t_5$, $i_{RS1}$ drops below $i_{CS1}$, causing a direction reversal of $i_{DC}$.

During ($t_6$-$t_8$), $C_{gs,S1}$ stops supplying charge and is instead charged by $i_{g,S1}$, which supplies charge to both $C_{gs,S1}$ and $C_{gd,S1}$. Notably, the mid-point voltage remains below $V_{DC}$ (by **P4**; considering influence of DC source) before $t_7$ keeping $i_{RS1}$ positive; after $t_7$, the mid-point voltage exceeds $V_{DC}$ (by **P4**; considering influence of DC source), reversing $I_{RS1}$'s direction. Meantime, the $C_{S2}$-conducted share of $i_L$ gradually commutates to $S_1$ (by **P4**; considering influence of $i_L$). At $t_8$, the current commutation is completed and $C_{S2}$ is fully charged, raising the mid-point voltage to $V_{DC}+R_{S1}(t_8)i_L(t_8)$. After $t_8$, as $v_{gs,S1}$ further increases, $R_{S1}$ decreases slowly (by **P3**), causing a slight drop in mid-point voltage and consequently a minor discharge of $C_{S1}$ and $C_{S2}$ via $R_{S1}$.

## Data availability

The data presented in this study are available in this manuscript.

## Code availability

No custom code was used in this study.

## Competing Interests Statement

The authors declare no competing interests.

## Supplementary Information

N/A